%% file: creativity.tex
\documentclass[manuscript]{acmart}

\AtBeginDocument{%
  }

\usepackage[most]{tcolorbox}
\usepackage{multirow}
\usepackage{makecell} 

\setcopyright{cc}
\setcctype{by}

\acmConference[CHI '27]{Proceedings of the ACM Conference on Human Factors in Computing Systems}{May 10--May 14, 2027}{Pittsburgh, PA, USA}
\acmBooktitle{Proceedings of the ACM Conference on Human Factors in Computing Systems (CHI '27), May 10--May 14, 2027, Pittsburgh, PA, USA}
\copyrightyear{2027} 
\acmYear{2027}

\usepackage{fontawesome}
\usepackage{multirow}
\usepackage[svgnames,table]{xcolor}
\input{sections/macros}

\makeatletter
\@ifpackageloaded{arydshln}{
   \newcommand{\dashedline}[1]{
    \cdashline{#1}[.3pt/1pt]\noalign{\vskip 1pt}
  }
}{
  \newcommand{\dashedline}[1]{
    \cmidrule[.00025mm]{#1}
  }
}
\makeatother
\begin{document}


\title{Characterizing Creativity in Data Visualization: Reflections and Future Directions}

\author{Tianwei Ma}
\email{tianwei-ma@uiowa.edu}
\orcid{0009-0009-9819-3982}
\affiliation{%
  \institution{University of Iowa}
  \city{Iowa City}
  \country{USA}
}

\author{Zinat Ara}
\email{zara@gmu.edu}
\orcid{0009-0003-2479-8846}
\author{Safwat Ali Khan}
\email{skhan89@gmu.edu}
\orcid{0000-0001-8220-4477}
\affiliation{%
  \institution{George Mason University}
  \city{Fairfax}
  \country{USA}
}

\author{Fanny Chevalier}
\email{fanny@cs.toronto.edu}
\orcid{0000-0002-5585-7971}
\affiliation{%
  \institution{University of Toronto}
  \city{Toronto}
  \country{Canada}
}

\author{Niklas Elmqvist}
\email{elm@cs.au.dk}
\orcid{0000-0001-5805-5301}
\affiliation{%
 \institution{Aarhus University}
 \city{Aarhus}
 \country{Denmark}
}

\author{Naimul Hoque}
\email{naimul-hoque@uiowa.edu}
\orcid{0000-0003-0878-501X}
\affiliation{%
  \institution{University of Iowa}
  \city{Iowa City}
  \country{USA}
}

\renewcommand{\shortauthors}{Ma et al.}

\newcommand{\fanny}[1]{\textcolor{purple}{[FC]: #1}}
\newcommand{\niklas}[1]{\textcolor{orange}{[NE]: #1}}
\setlength{\abovecaptionskip}{3pt}
\setlength{\belowcaptionskip}{-2pt}

\definecolor{anti-flashwhite}{rgb}{0.95, 0.95, 0.96}
\definecolor{amber}{rgb}{1.0, 0.49, 0.0}

\definecolor{myblue}{rgb}{0.2745,0.5098,0.7059}

\definecolor{myblue2}{rgb}{0.0,0.0,0.50}


\begin{abstract}
Creativity is valued in data visualization design and research, yet how it manifests remains unclear. Understanding creativity is essential for developing expressive visualization tools and representations. In this paper, we present two complementary studies to characterize creativity in data visualization.
First, a systematic review of 104 papers yields a \textbf{design space} spanning four themes:
design frameworks incorporating divergent and convergent thinking activities; creative representations developing unorthodox visualizations; visualization-enabled creativity support tools supporting creative tasks (e.g., writing); and understanding creativity, creative processes, and influencing factors. Second, qualitative interviews with 11 visualization practitioners and researchers explore practical challenges and contrast them with current academic framing through our design space.
Findings indicate that \textbf{artifacts or final products} are often disproportionately considered as the primary indicator of creativity, whereas the \textbf{design process} remains undervalued in practical. We conclude by presenting a definition of creativity in data visualization and directions for future research.
\end{abstract}

\ccsdesc[500]{Human-centered computing~Visualization theory, concepts and paradigms}

\keywords{Creativity, design, data visualization, literature survey.}


\begin{teaserfigure}
\centering
  \includegraphics[width=0.98\textwidth]{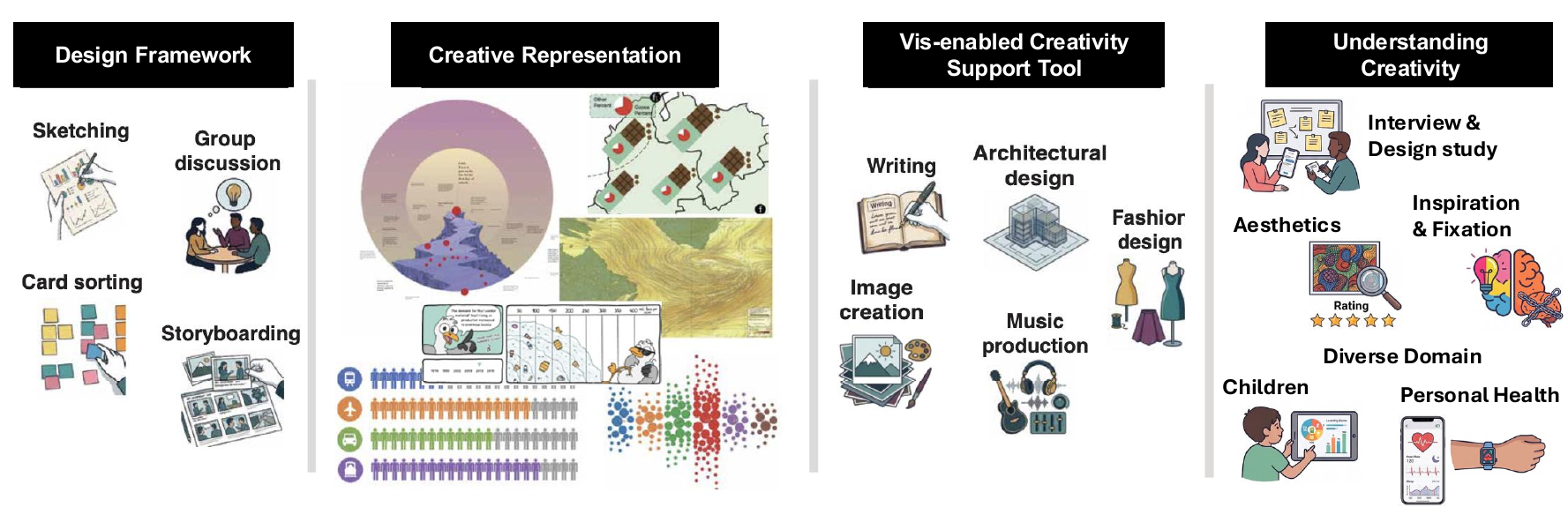}
  \caption{\textbf{Four broad creativity themes of the design space.}
  \textbf{Design framework} involving activities such as sketching, group discussion, and card sorting can foster human creativity regardless of the resulting visualization.
  \textbf{Creative representations} such as infographics, pictorials, and data comics can promote creativity through unusual layouts and personalized glyphs and icons.
  These representations typically rely on author-driven creative activities such as designing custom glyphs and selecting image segments for pictorials.
  \textbf{Vis-enabled Creativity Support Tools} designed with data visualization can support creative tasks such as writing, music production, fashion design, and image creation. 
  \textbf{Understanding Creativity} covers creative practices, empirical or theoretical research, and creative factors in visualization-related contexts. Icons in this figure were generated with Gemini.
  }
  \Description{The figure has four columns. Each column has a few pictures, icons, and texts.}
  \label{fig:teaser}
\end{teaserfigure}



\maketitle

\input{sections/1.introductions}

\input{sections/2.related_works}

\input{sections/3.method}
\input{sections/4.0.design_space}

\input{sections/5.interviews}
\input{sections/6.definition}

\input{sections/7.discussion}
\input{sections/8.conclusion}

\begin{acks}
    This work was supported partly by Villum Investigator grant VL-54492 by Villum Fonden.
    Any opinions, findings, and conclusions expressed in this material are those of the authors and do not necessarily reflect the views of the respective funding agencies.
\end{acks}

\bibliographystyle{ACM-Reference-Format}
\bibliography{creativity}

\end{document}

%% file: sections/macros.tex
\newboolean{showcomments}
\setboolean{showcomments}{true}

\ifthenelse{\boolean{showcomments}}
  {\newcommand{\nb}[2]{
    \fbox{\bfseries\sffamily\scriptsize#1}
    {\sf\small$\blacktriangleright$\textit{#2}$\blacktriangleleft$}
   }
   
  }
  {\newcommand{\nb}[2]{}
   
  }

%% file: sections/1.introductions.tex
\section{Introduction}

Creativity is typically tied to the emergence of new ideas~\cite{DBLP:conf/candc/Bonnardel99, sternberg1999handbook}.
It is an abstract construct and is often dubbed a puzzle, paradox, or mystery~\cite{boden2004creative}.
Indeed, researchers have come up with different theories and frameworks to model creativity.
Some propose that creativity stems from sudden breakthroughs and divergent thinking, while others relate creativity with a structural process of studying previous work and using methodical techniques to explore the possible solutions exhaustively~\cite{boden2004creative, gardner2011creating, mayer1992thinking}. 
Regardless of the underlying definition, ever since Shneiderman~\cite{DBLP:journals/tochi/Shneiderman00, DBLP:journals/cacm/Shneiderman07a} and Fischer~\cite{DBLP:conf/pdc/Fischer04} pointed out that computers have the potential to support and enhance human creativity, researchers have proposed numerous creativity support tools~\cite{DBLP:conf/chi/FrichV0BD19, DBLP:conf/ACMdis/FrichBD18}.
These tools are diverse in terms of tasks and user groups, ranging from tools to support video production and creative writing to sketching and design~\cite{DBLP:conf/ACMdis/FrichBD18}. 


Deriving ways to visualize data can be interpreted as a creative task~\cite{DBLP:journals/cacm/Shneiderman07a}.
M{\'{e}}ndez et al.~\cite{DBLP:conf/chi/MendezHN17} found that lower-level visualization design steps, such as the specific mapping of data points to visuals, can foster human creativity.
Similarly, Goodwin et al.~\cite{DBLP:journals/tvcg/GoodwinDJDDDKSW13} suggested that lower-level design activities (e.g., collaborative workshops) promote creativity in the early stages of the visualization design process. 
Another set of works focused on unorthodox or artistic representations of data, often through infographics~\cite{DBLP:journals/cgf/CoelhoM20, DBLP:journals/tvcg/KimSLDLPP17}, unusual layouts~\cite{DBLP:journals/tvcg/OffenwangerBCT24}, and by leveraging free-form drawings for glyphs and icons~\cite{xia2018dataink}.
Despite these efforts and growing interest, \textit{creativity remains an enigma} in visualization research, often craved by the community without any unified understanding of it. We have limited knowledge about how previous research conceptualized ``creativity'' and gaps that can be addressed to develop creative visualization tools and representations.
Towards this end, in this paper we conduct a literature survey to develop a \textbf{design space} that characterizes creativity in visualization design.
We analyzed 104 visualization papers that explicitly pursued creativity as a contribution, design goal, desired feature, or evaluation criteria. The resulting design space consists of four themes: design frameworks, creative representations, visualization-enabled creativity support tools, and understanding creativity (Figure~\ref{fig:teaser}).
Our findings suggest that prior research predominantly used unusual layouts and personalized glyphs and icons in forms of pictorials, infographics, and data comics to define \textbf{creative data visualization}~\cite{DBLP:journals/tvcg/YingSDYTYW23, 10.1145/3290605.3300335, DBLP:journals/cgf/TyagiZPKM22}.
However, creative data visualization does not necessarily require a novel or artistic representation; even a simple bar chart can be considered creative.
The design process that facilitates divergent and convergent thinking is often the key source of creativity in visualization design~\cite{DBLP:journals/tvcg/GoodwinDJDDDKSW13, DBLP:journals/tvcg/RobertsHR16, DBLP:journals/tvcg/RobertsRJH18}.
Even representations identified as ``creative'' depend predominantly on author-driven direct manipulation actions~\cite{direct_manipulation, direct_manipulation2}, such as mapping individual data points to marks, designing custom glyphs, and free-form drawing~\cite{xia2018dataink, DBLP:journals/tvcg/OffenwangerBCT24}.
We also noticed that algorithms described in our corpus focused primarily on helping the authoring process, while advanced AI or machine learning (ML) models appeared in only 14 of the 52 creative representation papers~\cite{DBLP:journals/tvcg/CuiWHWLZZ22}.
In addition, we found that data visualization can support the design of creativity support tools for a wide variety of tasks (e.g., writing~\cite{DBLP:conf/chi/HoqueMGSCKE24} and fashion design~\cite{DBLP:conf/chi/JeonJSH21}).
Finally, we identified gaps across themes, such as the limited understanding of students' and children's creative processes and the lack of consensus on how to evaluate creativity.

To understand how these findings relate to real-world practice, we further conducted qualitative interviews with 11 visualization practitioners and researchers, using the design space as a probe for discussion. The participants appreciated that our design space identified both the \textbf{design process} and the \textbf{artifacts} (final visual representations) as indicators of creativity. However, our findings indicate that in practical and organizational contexts, people often appreciate the final artifact as the creative outcome, but undervalue the design process and overlook the organizational constraints (e.g., short deadlines) under which it takes place.

In summary, this paper makes the following contributions:
\begin{itemize}
    \item A design space of creativity in data visualization, derived from a systematic review of 104 papers, that characterizes four themes and their dimensions and reveals gaps across themes.
    \item Findings from interviews with 11 visualization practitioners and researchers on how creativity manifests in practice and how it contrasts with its academic framing.
    \item A working definition of creativity in data visualization, together with design implications and directions for future research.
\end{itemize}


%% file: sections/2.related_works.tex
\section{Background and Related Work}
\label{sec:related_works}

The study of creativity in visualization sits at the intersection of theories and frameworks for creativity, HCI, and data visualization.
We provide a brief overview of these domains below in relevance to our work.

\subsection{Definition and Frameworks for Creativity}
\label{sec:def_creativity}

Creativity has been defined in multiple ways across disciplines.
Boden defines \textit{creativity} as \textit{``the ability to come up with ideas or artefacts that are }new\textit{, }surprising\textit{, and }valuable\textit{.''}~\cite[p.\ 1]{boden2004creative}
At idea level, Dean et al.\ operationalize a \textit{creative idea} 
as one that is simultaneously \textit{novel}, \textit{workable}, and \textit{relevant}---that is, it must be original, practically implementable, and applicable to the problem at hand~\cite{DBLP:journals/jais/DeanHRS06}.
This product-focused view reflects one strand within the broader four-P framework proposed by Rhodes~\cite{rhodes1961analysis}, who argued that creativity can be analyzed along four interrelated dimensions: Persons, Process, Press, and Products.
Beyond these definitional frameworks, scholars have also characterized creativity through three broad perspectives~\cite{sternberg1999handbook, DBLP:journals/tochi/Shneiderman00}. 
\textbf{Inspirationalists} emphasize ``Eureka'' moments,\footnote{A sudden flash of insight; the exclamation is traditionally attributed to Ancient Greek mathematician and inventor Archimedes.} while acknowledging that preparation and effort precede breakthroughs.
They encourage \textit{divergent thinking}~\cite{DBLP:conf/chi/FrichNHD21}---the spontaneous generation of multiple, varied and novel ideas by exploring a broad solution space from a given problem through techniques such as brainstorming, sketching, and concept mapping, in contrast to \textit{convergent thinking}~\cite{DBLP:conf/chi/FrichNHD21} which follows a particular set of logical steps including analytical evaluating, selecting, refining, and organizing these ideas to arrive at the most appropriate or effective solution~\cite{Guilford1950}.
\textbf{Structuralists} stress systematic exploration of possibilities, often through literature reviews, simulations, and iterative problem-solving steps such as planning, execution, and reflection~\cite{DBLP:journals/tochi/Shneiderman00}.
\textbf{Situationalists} view creativity as socially and environmentally situated, shaped by motivations, relationships, and peer recognition~\cite{sawyer2017group}.
They highlight collaborative practices such as participatory design and peer review.
This paper situates visualization design within these perspectives to identify gaps in the literature and guide future research.

\subsection{Creativity in Visualization Research}
\label{sec:viz_creativity}

The application of creativity in visualization design has gained momentum in recent years.
Early discussions of creativity in visualization often centered on the tension between artistic expression and analytical rigor.
Vande Moere and Purchase~\cite{DBLP:journals/ivs/MoereP11} highlighted the creative aspects of information visualization, arguing that aesthetic considerations play a crucial role in effective visualization design.
Scholtz~\cite{DBLP:conf/ieeevast/Scholtz06} proposed creativity as an area of evaluation for visualization systems.
She further proposed metrics---such as quality of solutions, number of unique solutions explored, serendipitous solutions, satisfaction with solutions, and more---as metrics for evaluating creative aspects of a visualization system.
Parsons et al.~\cite{DBLP:conf/visualization/ParsonsSP21} interviewed 15 visualization practitioners to understand \textit{design fixation}: designers' tendency to adhere blindly or prematurely to a set of ideas that limit creative outcomes. 

Beyond theoretical discussion on the importance of creativity in visualization, research on creativity in visualization falls into three broad directions.
The first direction proposes design frameworks and methodologies to support creative practices in visualization design~\cite{howard2008describing, DBLP:conf/avi/BigelowDFM14, DBLP:conf/ACMdis/BressaWKWV19, DBLP:journals/tvcg/GoodwinDJDDDKSW13, DBLP:conf/chi/MendezHN17, DBLP:journals/tvcg/RobertsHR16}.
This direction can be called a representation-agnostic approach since frameworks in this category do not explicitly propose any representations; rather, the frameworks involve design practices (e.g., discussion, sketching, card sorting, etc.) for promoting creativity.
For example, Kerzner et al.~\cite{DBLP:journals/tvcg/KerznerGDJM19} proposed ``Creative Visualization-Opportunities Workshop,'' a design workshop that focuses on using design activities such as discussion and sketching to identify requirements for a visualization project.
The workshop emphasizes bringing different stakeholders in a common workshop for assessing the feasibility of the future project.
The Explanatory Visualization Framework~\cite{DBLP:journals/tvcg/RobertsRJH18} encourages students' creative thinking by asking them to manually explore different visual mappings between data and primitive shapes.

The second direction focuses on developing novel visual representations.
These representations typically rely on unorthodox layouts and artistic expressions through free-form drawings, glyphs, and icons. 
For example, DataInk~\cite{xia2018dataink} offers a direct manipulation interface to draw and organize glyphs and bind data to produce ``expressive'' and ``creative'' visual representations.
Similarly, the \textit{visual sedimentation} technique takes inspiration from the physical phenomenon of sedimentation and then uses the animated movement of data units to visualize the streaming data~\cite{DBLP:journals/tvcg/HuronVF13}.
The authors claim representations created with such animation as creative. The final direction is the use of visualization to develop Creativity Support Tools (CSTs). For example, HaLLMark~\cite{DBLP:conf/chi/HoqueMGSCKE24} helps writers track their interactions with an LLM during writing, and Visual Story-Writing~\cite{visual-storywriting} uses visual representations as input for AI-mediated writing of stories. 


The above research clearly shows interest towards integrating creativity in visualization design.
However, it also shows that the interest is quite diverse.
Thus, we believe that a unified understanding of this design space is timely, especially in light of emerging AI-powered visualization authoring tools. 
Parsons et al.~\cite{DBLP:conf/visualization/ParsonsSP21} found that one of the reasons designers fixate on a specific visualization is because current authoring tools do not offer creative exploration of different designs.
Our work outlines several design implications for integrating creativity in authoring tools.

It is worth acknowledging upfront that we have not considered works outside academic research in our analysis for developing the design space.
For example, data journalism in popular news outlets (e.g., the New York Times) has played a prominent role in devising creative data visualizations that engage readers.
Similarly, artists such as Giorgia Lupi, Nadieh Bremer, and Federica Fragapane have successfully merged data visualization with art.
Venues such as the VIS Arts Program (VISAP) have attracted a great deal of attention in the community. While we do not include these works in the design space, the feedback from the representative communities is incorporated through the interviews with journalists, designers, and researchers.


%% file: sections/3.method.tex
\begin{figure}
    \centering
    \includegraphics[width=0.95\linewidth]{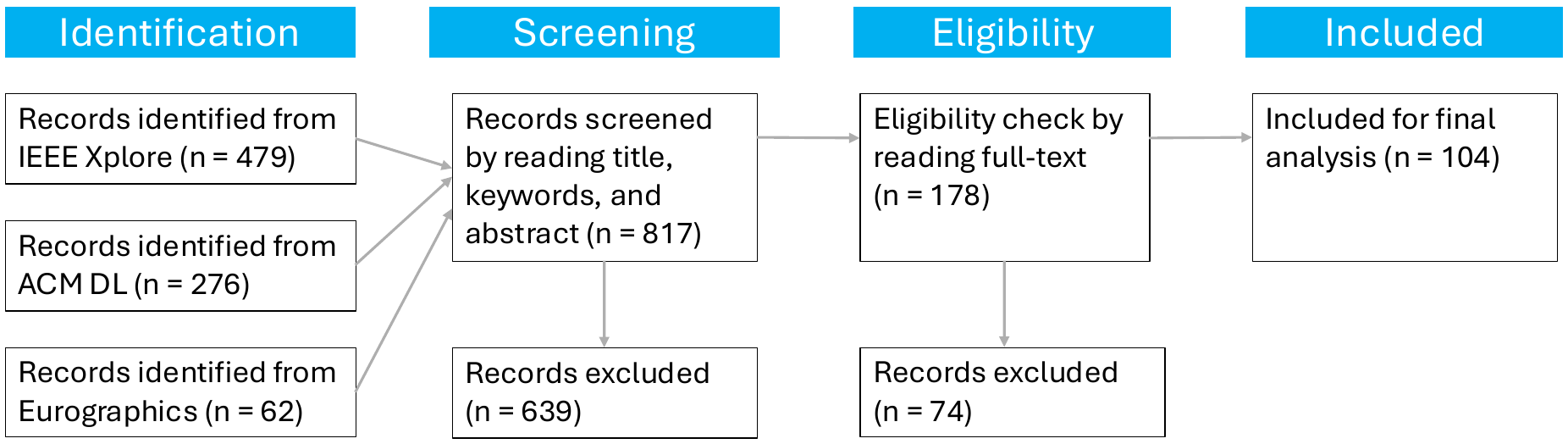}
    \caption{\textbf{Paper collection steps based on the PRISMA~\cite{Pagen71} method for literature review.} We \textit{identified} 817 papers initially through keyword search.  We \textit{screened} these papers by reading their titles, abstracts, and keywords, and excluded ones that were clearly out of scope for the survey. We then checked \textit{eligibility} by reading the full-text articles and included papers that were about visualization and discussed creativity as a repeated theme. We \textit{included} 104 papers for final review.}
    \Description{A four-stage PRISMA flowchart: three database sources feed into screening, eligibility, and inclusion, with excluded records branching down.}
    \label{fig:method}
\end{figure}

\begin{figure}
    \centering
    \includegraphics[width=0.9\linewidth]{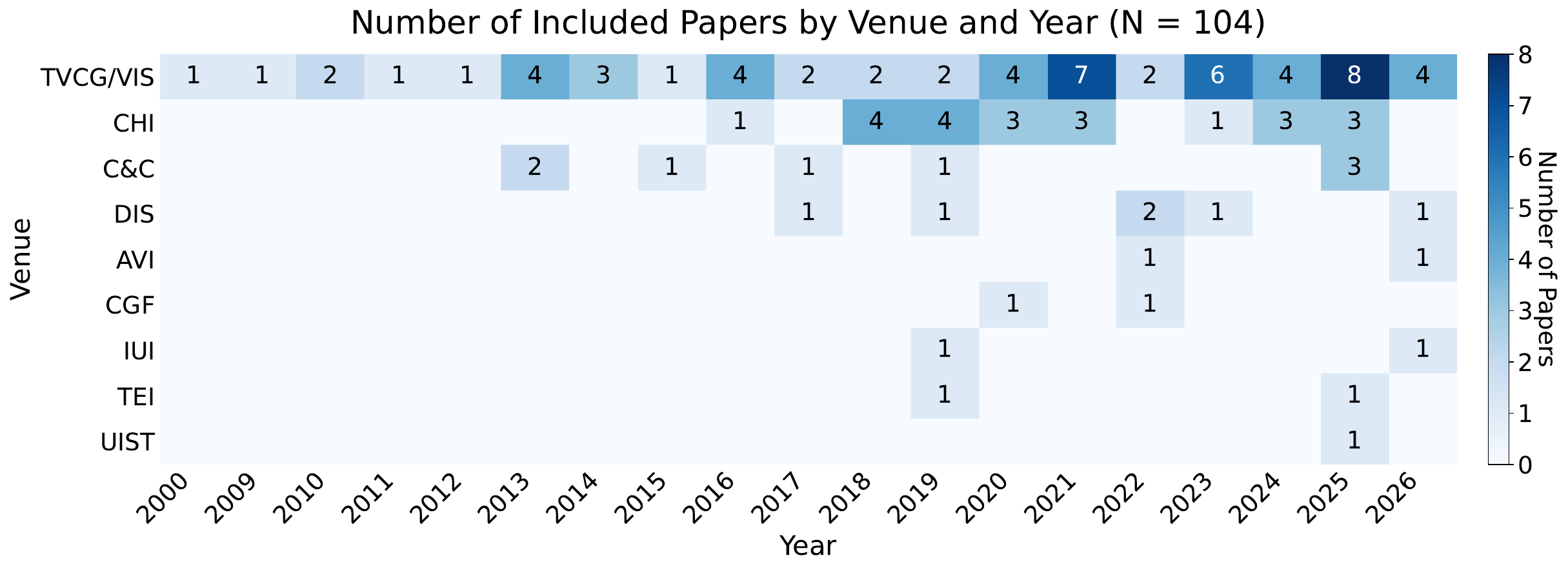}
    \caption{\textbf{Papers included for analysis.} Papers appearing in InfoVis, SciVis, VAST, VIS, and TVCG are merged together.} 
    \Description{A heatmap with venues as rows and years as columns, where darker cells mean more included papers.}
    \label{fig:heatmap}
\end{figure}

\section{Design Space Methodology}

We conducted a systematic review of prior research to characterize creativity in visualization research. We adopted the PRISMA 2020 (Preferred Reporting Items for Systematic Reviews and Meta-Analyses)~\cite{Pagen71} process for conducting the survey. The PRISMA method typically involves four steps: 1) Identification; 2) Screening; 3) Eligibility; and 4) Included. We followed these steps to collect the papers (\autoref{fig:method}). In the following, we describe each of the steps.

\subsection{Identification}
We identified potential papers for the survey at this stage. We focused on digital libraries and academic venues relevant to visualization, HCI, and creativity research. The venues are:

\begin{itemize}
    \item IEEE Xplore: VIS, TVCG
    \item ACM DL: CHI, UIST, DIS, IUI, CSCW, C\&C, AVI, TEI 
    \item EuroGraphics DL: EuroVis, CGF
\end{itemize}

We used a wide range of keywords for searching in the libraries. Previous research on creativity in HCI and AI typically focused on using keywords such as \textit{creativity} or \textit{creative}~\cite{DBLP:conf/chi/FrichV0BD19, DBLP:conf/ACMdis/FrichBD18}. However, we decided to expand the keywords to ensure high recall of the relevant papers. The specific search query was: \textit{[(visualiz* OR visual analytics) AND (creativ* OR aesthetic OR expressive OR artistic OR brainstorm* OR idea* OR inspiration OR fixation OR convergent OR divergent)]}. The initial search resulted in 479 papers from IEEE, 276 papers from ACM, and 62 papers from EuroGraphics.

\subsection{Screening}
The aim of this stage is to filter out papers that are not \textit{obviously} relevant to this survey. This includes papers that are not about data visualization or papers that clearly focus on the analytical side of data visualization. One author of this work manually read the titles, abstracts, and keywords of the collected papers and identified these papers. A second author read the papers (titles, abstracts, and keywords) flagged by the first author as boundary cases to reach decisions.  We retained a total of 178 papers after this stage.

\subsection{Eligibility}
This is the final stage to identify relevant papers. Since we filtered out all non-visualization papers before this stage, we focused on identifying whether a paper has a significant creativity angle to it. One author read the full papers and retained papers where creativity appeared as a recurring theme in the paper rather than a mere adjective or filler word. This criterion is particularly important, as it enforces that the authors of a paper  consciously pursue creativity as a desired goal, feature, or contribution. We note that this is a relatively strict criterion since, conceptually, some papers may not pursue creativity explicitly but contain a component that can be considered creative. However, without this criterion, the survey will lack an objective inclusion policy, reproducibility, and focused results. A second author read full papers flagged by the first coder as boundary cases to reach decisions. Papers were also discussed in the group meetings to decide eligibility. The final set included 104 papers. Figure~\ref{fig:heatmap} shows the distribution of the papers.

\subsection{Included}
We analyzed the collected papers at this stage. We adopted a grounded theory approach~\cite{glaser2017discovery}.
At first, three authors of this paper independently read and open-coded 10 randomly chosen papers.
The authors then met several times to discuss the dimensions for a codebook.
This resulted in the initial codebook for our analysis. 

At this stage, we created a new set of papers with 5 papers from the initial set and 5 new randomly chosen papers.
The coders, including a fourth author who did not participate in the initial coding process, then independently coded this set of papers following the codebook.
The inter-annotator agreement after this stage was 0.85 (Jaccard's similarity).
We resolved disagreements through discussion and finalized the codebook.
After that, the papers were equally divided among the three initial coders. The key themes and the complete codebook, including dimensions, codes, and definitions, are presented in the following section. The coded corpus of all 104 papers is available in the \textbf{supplemental materials}. 

\definecolor{lightblue}{RGB}{240,248,255} 

\begin{table*}[t]
    \centering
    \caption{\textbf{
    The four types of creativity papers identified in our analysis.} The numbers indicate their frequency in the corpus. The themes are not mutually exclusive.}
    \Description{A three-column table listing the four creativity themes, with each theme's paper count, description, and example references.}
    \label{tab:type_creativity}
    \rowcolors{2}{White}{lightblue}
    \begin{tabular}{p{3cm}p{8cm}p{1.5cm}}
    \toprule
     
    \textbf{Type} & \textbf{Description} & \textbf{Examples} \\ 
    \toprule

    \textbf{Design Framework (21)} &  
    Research and design methods with design activities (e.g., collaboration, 
    ideation, and data collection) focusing on promoting creativity in different stages of a visualization 
    project. & 
    \cite{DBLP:journals/tvcg/KerznerGDJM19} 
    \cite{DBLP:journals/tvcg/RobertsRJH18} \\

    \textbf{Creative Representation (52)} &  
    Visual representations that are claimed to be creative in the papers. 
    The representation can rely on user interactions and/or computational 
    methods. & 
    \cite{DBLP:journals/tvcg/OffenwangerBCT24} 
    \cite{DBLP:conf/chi/RomatRHDAH20} \\

    \textbf{Vis-enabled Creativity Support Tool (28)} & 
    Systems or methods that utilize data visualization to support a creative 
    task other than visualization design (e.g., writing, fashion design, 
    music production, etc.). & 
    \cite{DBLP:conf/chi/HoqueMGSCKE24} 
    \cite{DBLP:conf/chi/JeonJSH21} \\

    \textbf{Understanding Creativity (17)} & 
    Empirical or theoretical research that investigates creativity, creative 
    practices, or factors that influence creativity in visualization-related 
    contexts. These studies aim to understand how creativity occurs, rather 
    than primarily proposing a method or system to support it. & \cite{DBLP:conf/chi/BaigelenovSP25} \cite{DBLP:journals/tvcg/BakoLBL23}
    \\

    \bottomrule
    \end{tabular}
\end{table*}

%% file: sections/4.0.design_space.tex
\begin{figure*}
    \centering
    \includegraphics[width=0.95\textwidth]{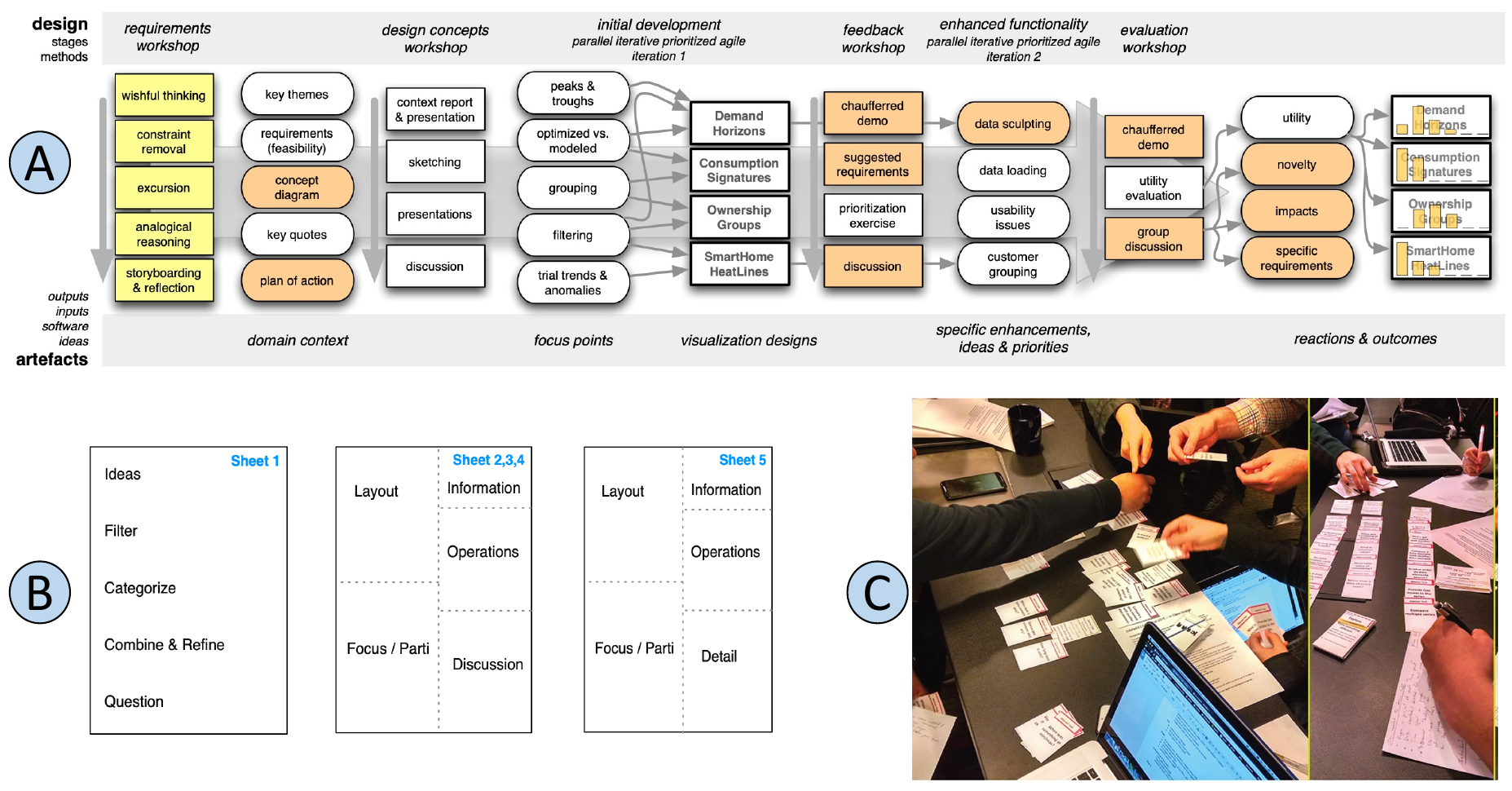}
    \caption{\textbf{Example creative design frameworks.} 
    (A) The workshop-based framework proposed by Goodwin et al.~\cite{DBLP:journals/tvcg/GoodwinDJDDDKSW13} for energy analysts to identify visualization requirements.
    The authors explicitly integrated design activities (yellow rectangles on the left) such as wishful thinking, excursion, and storyboarding to promote creativity.
    Here rectangles are techniques and round-edged rectangles are concepts.
    Orange colors indicate techniques and concepts where creativity was strong among the energy analysts.
    (B) The Five Design Sheet (FdS)~\cite{DBLP:journals/tvcg/RobertsHR16} divided the visualization design process into five sheets where sheet 2, 3, and 4 are identical.
    Designers will sequentially sketch and fill out the sheets to come up with the final design.
    (C) VIZITCARDS~\cite{DBLP:journals/tvcg/HeA17} proposes different kinds of cards (e.g., domain, abstract, layout, and inspiration cards) that students can use to collaboratively brainstorm about a visualization design.}
    \Description{This figure has three subfigures. The first subfigure shows a flowchart of interconnected rectangles. The second subfigure contains three rectangles. The third subfigure shows a few people working with some sticky notes.}
    \label{fig:example_design_frameworks}
\end{figure*}

\section{Design Space}
\label{sec:results}

This section presents the design space based on our analysis. The design space consists of
 four broad themes: \textbf{Design Frameworks, Creative Representations,}  \textbf{Visualization-enabled Creativity Support Tools (CST)}, and \textbf{Understanding Creativity}.
Table~\ref{tab:type_creativity} provides a formal definition for each of the four themes along with examples. Among the 104 papers, \textit{design frameworks} appeared in 21 papers, \textit{creative representations} in 52 papers, \textit{vis-enabled CSTs} in 28 papers, and \textit{understanding creativity} in 17 papers. 
Note that the themes are not mutually exclusive; a paper can fall under multiple themes of creativity. For example, Dear Pictograph~\cite{DBLP:conf/chi/RomatRHDAH20} falls under \textit{understanding creativity} since the paper wanted to understand the role of personalization and immersion in making pictographs more enjoyable and creative. However, the paper also presents an interface for authoring creative glyph-based and 3D pictographs. Thus, the paper falls under \textit{creative representation} too.  Design patterns for data comics~\cite{bach2018designpatterns} falls under both \textit{framework} and \textit{creative representation} since the paper proposed design patterns with specific narrative purposes and panels that designers can use to create data comics as well as presented new data comics created using the patterns. Below we organize the major findings based on the four creativity themes.

\input{sections/4.1.frameworks}
\input{sections/4.2.representation}
\input{sections/4.3.cst}
\input{sections/4.4.understanding_creativity}

\subsection{Key Insights and Observations}
We discuss key insights observed across the themes below.

\subsubsection{Process vs. Artifact}
We observed two primary purposes of the four identified themes. While themes with \textit{frameworks}  and \textit{CSTs} focus more on supporting design \textbf{processes}, \textit{creative representation} focuses on \textbf{artifacts} (i.e., data visualization) that are labeled as ``creative''. Frameworks and CSTs do produce artifacts, but they are not the main focus or labeled as ``creative''.
There is also a difference between frameworks and CSTs in terms of the process they support and the role data visualization plays in that process.
The framework theme focuses on design processes ``for'' creating data visualization, while the CST theme focuses on design processes ``with'' data visualization for creating other creative artifacts. The \textit{understanding creativity} theme studies both processes and artifacts in the other three themes.

\subsubsection{Epistemic Gap between Design Frameworks and Creative Representation}
\label{sec:gap}
We noticed an epistemic gap between design frameworks and creative representation themes in our review. The activities in the design frameworks and the authoring actions in the tools share similarities but are often discussed separately within the literature. For example, almost all design frameworks use sketching as a design activity~\cite{DBLP:journals/tvcg/GoodwinDJDDDKSW13, DBLP:journals/tvcg/RobertsHR16}. Similarly, several tools for developing creative representations use sketching as an authoring action~\cite{xia2018dataink, DBLP:journals/tvcg/OffenwangerBCT24}. Other activities such as data collection and binding were also common across the themes.
A stronger integration of the two themes might enable stronger creative explorations during early-stage conceptualization and development phases.

\subsubsection{Lack of Understanding about Children's and Students' Creative Process}
\label{sec:student}
We noticed another epistemic gap between design frameworks and understanding creativity themes. There were only 2 studies involving students~\cite{DBLP:journals/tvcg/BakoLBL23, iStoryline} and no papers with children in the understanding creativity theme. These counts are lower than the design framework theme, which had 4 papers with students and 2 papers with children. This indicates that while researchers have proposed several creative design frameworks for students and children, there is a lack of studies to  understand the design processes of the target users. This user group is free from predefined workflows; thus, it would be interesting to study their unbiased and natural design process for creating data visualization.

\subsubsection{No Consensus on Evaluation}
\label{sec:evl}
We found several evaluation strategies for creative representation without a clear consensus. Several papers used example galleries and demonstrations to show the creativity of the artifact~\cite{schroeder2015visualization, 10.1145/3290605.3300335, DBLP:conf/chi/ZhangSBC20}. Scholtz~\cite{DBLP:conf/ieeevast/Scholtz06} proposed quantitative measures such as the number of unique solutions explored, serendipitous solutions, satisfaction with solutions, etc. to measure the creative aspects of a visualization. He et al.~\cite{9903341} proposed BeauVis, a likert-scale-like rating scale with five items: ``enjoyable,'' ``likable,''
``pleasing,'' ``nice,'' and ``appealing'' to measure aesthetic pleasure associated with a visualization. Users will rate each item on a scale ranging from ``strongly disagree'' to ``strongly agree''. Overall, these are valuable contributions that should be synthesized for creating research guidelines.

%% file: sections/4.1.frameworks.tex
\begin{table}[t]
    \centering
    \caption{\textbf{Dimension, codes, and definition for Design Frameworks}}
    \Description{This is a table with three columns and five rows. Each row contains several nested rows.}
    \renewcommand{\arraystretch}{1.3}
    \begin{tabular}{m{2cm}p{6cm}m{6cm}}
    \toprule
     \textbf{Dimension}  & \textbf{Code} & \textbf{Definition}  \\
     \toprule
     \multirow{4}{=}{\centering Creativity Theory}
         & \multicolumn{2}{l}{\textit{Which theory of creativity is relevant to the design framework?}} \\ \dashedline{2-3}
         & Inspirationalist & Focuses on a sudden flash of inspiration and breakthrough. \\ 
         & Structuralist  & Focuses on systematic exploration of solutions through clear steps.\\ 
         & Situationalist  & Focuses on social and environmental factors behind creativity.  \\
         \hline
     \multirow{4}{=}{\centering Authoring Stage} & \multicolumn{2}{l}{\textit{Which stage of the design process does the framework target?}} \\ \dashedline{2-3}
         & Early-stage Conceptualization  & Focuses identifying scopes, requirements, and possible solutions for a project.\\ 
         & In-development  & Focuses on the active development of a representation.\\
         & Post-development  & Focuses on post-development reflection, critique, and revisions.\\
         \hline
    \multirow{10}{=}{\centering Design Activity}
    & \multicolumn{2}{l}{\textit{What design activities does the framework support/promote?}} \\ \dashedline{2-3}
    
    & Ideation/divergent thinking & Sketching, drawing, storyboarding, constraint removal, piloting methods, wishful thinking, and analogical reasoning. \\ 
    
    &Synthesis/convergent thinking & Card sorting, grouping elements, affinity diagramming, and storyboarding (also appears as an ideation mechanism). \\ 
    
    & Collaboration & Group discussion, excursion, and focus group. \\ 
    
    & Data Collection & Interviews, surveys, example search, raw data, and literature reviews. \\ 
    
    & Reflection & Writing, looking back and criticizing solutions.  \\ 
    \hline
    \multirow{3}{*}{Target Users} & \multicolumn{2}{l}{\textit{Who are the target users?}} \\ \dashedline{2-3}
    & Students, Children, Domain Experts, Visualization Experts, No specific criteria & Not applicable \\
    \hline

    Open-ended & Notes &Open-ended comments about the framework. \\
    \bottomrule
    \end{tabular}
    \label{tab:design_framework_codebook}
\end{table}

\subsection{Creativity in Design Frameworks}
\label{sec:rq1}

Articles presenting a design \textit{process} or \textit{method} for promoting creativity in a visualization project fall under this theme. 
Goodwin et al.~\cite{DBLP:journals/tvcg/GoodwinDJDDDKSW13} mentioned that \textit{``Establishing requirements can be considered a fundamentally creative process whereby requirements analysts and stakeholders work collaboratively to generate ideas for software systems.''}
These frameworks typically involve different types of creative design activities.
For instance, Goodwin et al.~\cite{DBLP:journals/tvcg/GoodwinDJDDDKSW13} deliberately integrated five creative design activities: wishful thinking, constraint removal, excursion, analogical reasoning, and storyboarding and reflection in their framework (\autoref{fig:example_design_frameworks}a).
The Five Design Sheet (FdS)~\cite{DBLP:journals/tvcg/RobertsHR16} approach introduces five structured sheets where visualization authors use sketching to brainstorm about core ideas, filters, and layouts, and write down discussion points and insights (\autoref{fig:example_design_frameworks}b). \autoref{fig:example_design_frameworks}c shows a design framework that is designed to educate university students about visualization design through different card sorting activities. This theme contains four dimensions (\autoref{tab:design_framework_codebook}). They are discussed below.

\subsubsection{Creativity Theory}
\label{sec:df_theory}
This dimension captures how the proposed framework relates to creativity theories from \autoref{sec:def_creativity}.
Nearly all frameworks (20 out of 21 papers) attempt to introduce \textit{structure} into visualization design and thus align well with the \textit{structuralist theories}. Although the frameworks focus on structure, all include activities that support divergent thinking (e.g., sketching), aligning highly with the inspirationalist theory. 
In contrast, situationalist approaches were less popular, appearing in 13 out of 21 papers. Collaborative activities were key indicators of the situationalist approach~\cite{DBLP:journals/tvcg/GoodwinDJDDDKSW13, DBLP:journals/tvcg/HeA17}. 10 papers combined all three approaches.


\subsubsection{Authoring Stage}
\label{sec:df_authoring}
This dimension identifies the stage of the authoring process that the frameworks aim to support. Almost all papers supported early-stage conceptualizations (20 out of 21 papers). This was expected as the purpose of frameworks was to enable systematic ways to conceptualize visualization design. In-development support was also common (17 out of 21 papers) as most frameworks provide some sort of tools with structured design or guidelines that can be used to create visualizations within the frameworks. For example, Construct-A-Vis~\cite{DBLP:journals/tvcg/BishopZPSRH20} provides a tool for children to manipulate visual elements. Interactive Data Comics~\cite{DBLP:journals/tvcg/WangRCRMB22} supports authors during construction of a data comic by helping them arrange, edit, and refine comic components as the artifact develops. There were 4 papers that did not provide any mechanism to develop the actual artifact; rather, they entirely focused on supporting ideation about the visualization~\cite{DBLP:journals/tvcg/KerznerGDJM19, DBLP:journals/tvcg/HeA17, DBLP:journals/tvcg/RobertsHR16, DBLP:conf/ACMdis/BressaWKWV19}. Post-development support was relatively less common, appearing in 10 papers out of 21.

\subsubsection{Creative Design Activity}
\label{sec:df_activity}
This dimension captures low-level design activities (sketching, card sorting, affinity diagramming, etc.) listed as part of a framework. This is the core element that the frameworks rely on to foster creativity in the design process. As presented in Table~\ref{tab:design_framework_codebook}, we combined the activities into five broad categories: \textit{ideation/divergent thinking (21), synthesis/convergent Thinking (19), Reflection (16), Collaboration (13),} and \textit{Data Collection (10)}. The reviewed frameworks provide a mix of divergent and convergent activities, with sketching being the most common activity. 

\subsubsection{Target Users}
\label{sec:df_user}
Non-experts and domain experts are the typical target users of the frameworks. 
We noticed that researchers have primarily developed the frameworks to improve engagement with non-experts and domain experts.
The frameworks work as vehicles for collecting requirements from non-experts and domain experts, and for helping them produce visual representations for their problems~\cite{DBLP:journals/tvcg/WoodBD14, DBLP:journals/tvcg/McKennaMAM14}.
Another major motivation is to educate non-experts and students about key visualization concepts such as marks, channels, tasks, color perception, and layouts~\cite{DBLP:journals/tvcg/BishopZPSRH20, DBLP:journals/tvcg/HeA17}.

%% file: sections/4.2.representation.tex
\begin{figure*}[t]
    \centering
    \includegraphics[width=\linewidth]{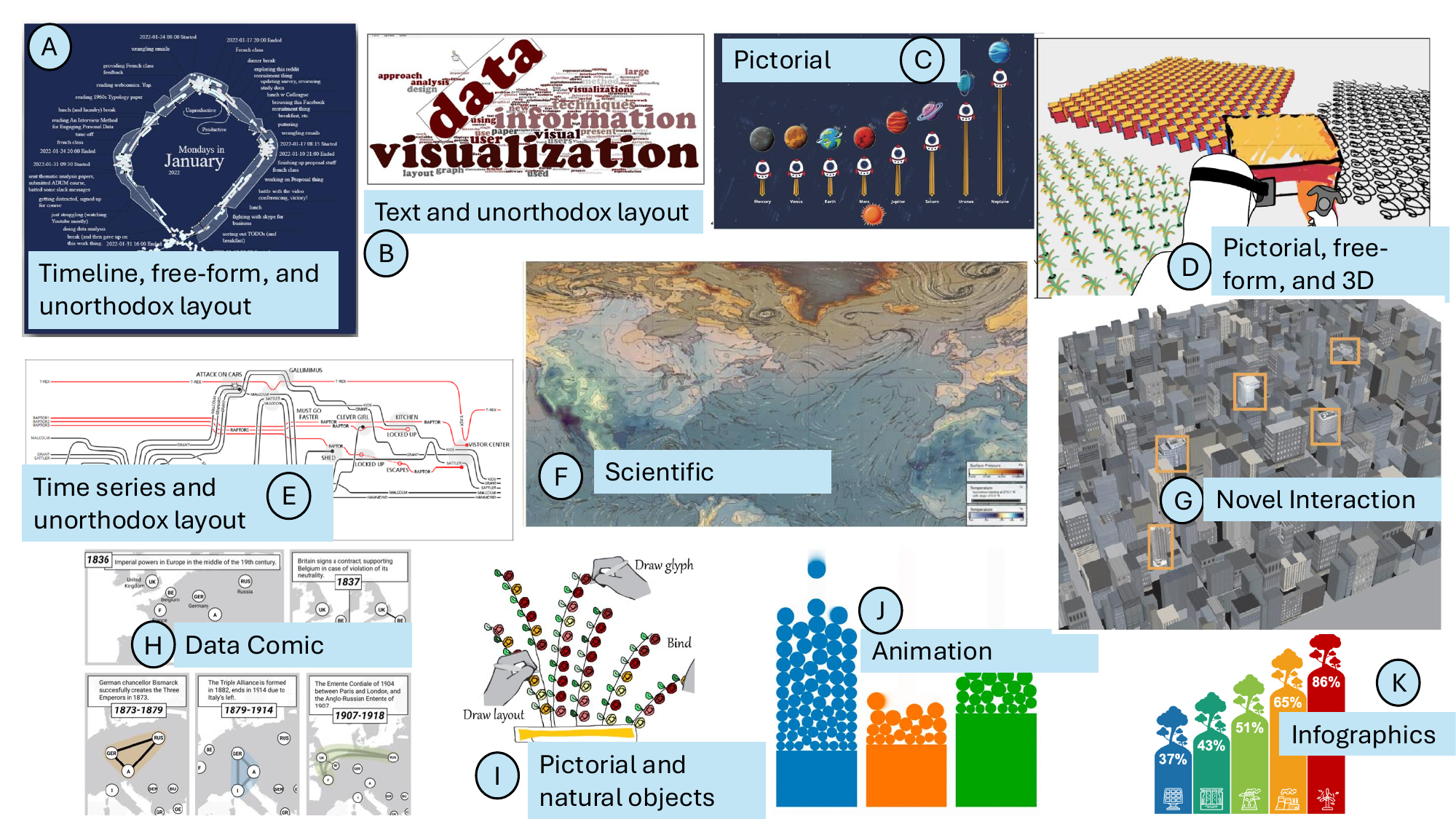}
    \caption{\textbf{Example creative representations with different properties. 
    }
    (A) TimeSplines~\cite{DBLP:journals/tvcg/OffenwangerBCT24} uses free-form sketch and unorthodox layouts to represent time series. (B) Word cloud~\cite{maniwordle} representation uses text and unorthodox layout.
    (C) DataQuilt~\cite{DBLP:conf/chi/ZhangSBC20} lets users extract visual elements from images to form pictorials.
    (D) Dear Pictograph~\cite{DBLP:conf/chi/RomatRHDAH20} uses free-form glyphs and 3D environment to produce pictorials.
    (E) Storyline visualization~\cite{iStoryline} representing the narrative of the Lord of the Rings.
    (F) Schroeder and Keefe~\cite{schroeder2015visualization} proposed a system where artists can paint on top of scientific visualization.
    (G) Custom focus and zooming interaction proposed by Qu et al.~\cite{DBLP:journals/tvcg/QuWCWC09} to investigate urban environments.
    (H) Graph comics by Bach et al.~\cite{DBLP:conf/chi/BachKHCKR16}.
    (I) Pictorials with natural objects (i.e., leaf) from DataInk~\cite{xia2018dataink}.
    (J) Inspired by the physical sedimentation process in rivers, Huron et al.~\cite{DBLP:journals/tvcg/HuronVF13} proposed an animation for streaming data.
    (K) An infographic example by Cui et al.~\cite{DBLP:journals/tvcg/CuiWHWLZZ22}.}
    \Description{This figure has 11 subfigures. Each subfigure shows a screenshot of a data visualization.}
    \label{fig:rep_examples}
\end{figure*}

\begin{table}[ht]
    \centering
    \caption{\textbf{Dimension, codes, and definition for Creative Representation}}
    \Description{This is a table with three columns and six rows. Each row contains several nested rows.}
    \renewcommand{\arraystretch}{1.2}

    \begin{tabular}{m{2cm}m{6cm}m{6cm}}
    \toprule
    \textbf{Dimension} & \textbf{Code} & \textbf{Definition} \\
    \toprule

    \multirow{4}{=}{\centering Representation Type}
        & \multicolumn{2}{l}{\textit{What is the base type of the representation?}} \\ 
        \dashedline{2-3}

        & Area, Bar, Circle, Diagram, Distribution, Grid and Matrix, Line, Map, Point, Table, Text, Trees and Networks, \textcolor{blue}{Unit viz, Scivis, No specific criteria} 
        & Taxonomy adopted from Borkin et al.~\cite{DBLP:journals/tvcg/BorkinVBISOP13}. Newly added options are highlighted in \textcolor{blue}{blue}.\\
    \hline

    \multirow{8}{=}{\centering Representation Properties} 
        & \multicolumn{2}{l}{\textit{What are the properties of the representation that make it creative?}} \\ 
        \dashedline{2-3}

        & 3D, Composite, Pictorial, Timeseries, Animation, Data Comic, Illustrative text, \textcolor{blue}{Free-form, AR/VR, Infographic, Scientific, Unorthodox layout, Natural objects, New interaction}
        & Also adopted from the representation properties proposed by Borkin et al.~\cite{DBLP:journals/tvcg/BorkinVBISOP13}. Newly added options are highlighted in \textcolor{blue}{blue}.\\

    \hline

    \multirow{12}{=}{\centering Authoring Actions}
        & \multicolumn{2}{l}{\textit{What actions can an author perform to create a creative representation?}} \\ 
        \dashedline{2-3}

        & Draw/sketching 
        & Authors can freely draw and sketch shapes and texts in a digital screen. \\

        & Common actions 
        & Authors can apply common data visualization actions such as select, add, remove, etc. to create/refine a representation.\\

        & Data binding 
        & Authors can control the binding of data to specific marks and channels.\\

        & Annotate 
        & Authors can freely write text and explain data points. \\

        & Re-purposing 
        & Authors can re-purpose and integrate existing image and shape into new visualization.\\
    \hline

    \multirow{3}{=}{\centering Manipulable Objects} 
        & \multicolumn{2}{l}{\textit{What objects can authors manipulate that enhance creativity and expressiveness?}} \\ 
        \dashedline{2-3}

        & Mark, Icon, Text, Color, Image, Free-form object, (Scales, Axis, Labels), Path, Link, Layout
        & Visualization objects that authors can manipulate. The objects were inspired by the taxonomy proposed by Satyanarayan et al.~\cite{DBLP:journals/tvcg/SatyanarayanLRH20}.\\
    \hline

    \multirow{8}{=}{\centering Computational Support} 
        & \multicolumn{2}{l}{\textit{What kind of computational supports are key behind the representation?}} \\ 
        \dashedline{2-3}

        & Algorithms 
        & The paper uses advanced algorithms and heuristics for creating representations. Common automation methods such as automatically detecting attribute types are not included, as they are not relevant to the topic.\\

        & AI/ML 
        & Uses data-driven AI/ML models and techniques.\\
    \hline

    Study Type 
    & Interview, Survey, Gallery, Reproduction, Free-form, Design Study, 
    Technical Evaluation, Usability Testing, Case Study, Controlled Study, 
    Field Study, Reading Study, Other
    & The empirical method used to evaluate or investigate the creative representation. \\

    Open-ended 
        & Notes 
        & Open-ended comments about the representation.\\

    \bottomrule
    \end{tabular}

    \label{tab:representation_codebook}
\end{table}

\subsection{Creative Representations}
\label{sec:rq2}

This theme summarizes the development of creative visual representations.
Works here emphasize the atypical, whimsical, or unorthodox nature of the visual representations proposed in the system.
\autoref{fig:rep_examples} presents a few example representations that we found in our analysis.
\autoref{tab:representation_codebook} presents the relevant dimensions, codes and definitions for creative representations. The dimensions and codes capture different aspects of the authoring process---from the representation type to creative properties of the representation, author-driven actions, objects that authors can manipulate, and computational methods that assist authors. Below, we briefly describe the dimensions.

\begin{figure}
    \centering
    \includegraphics[width=0.9\linewidth]{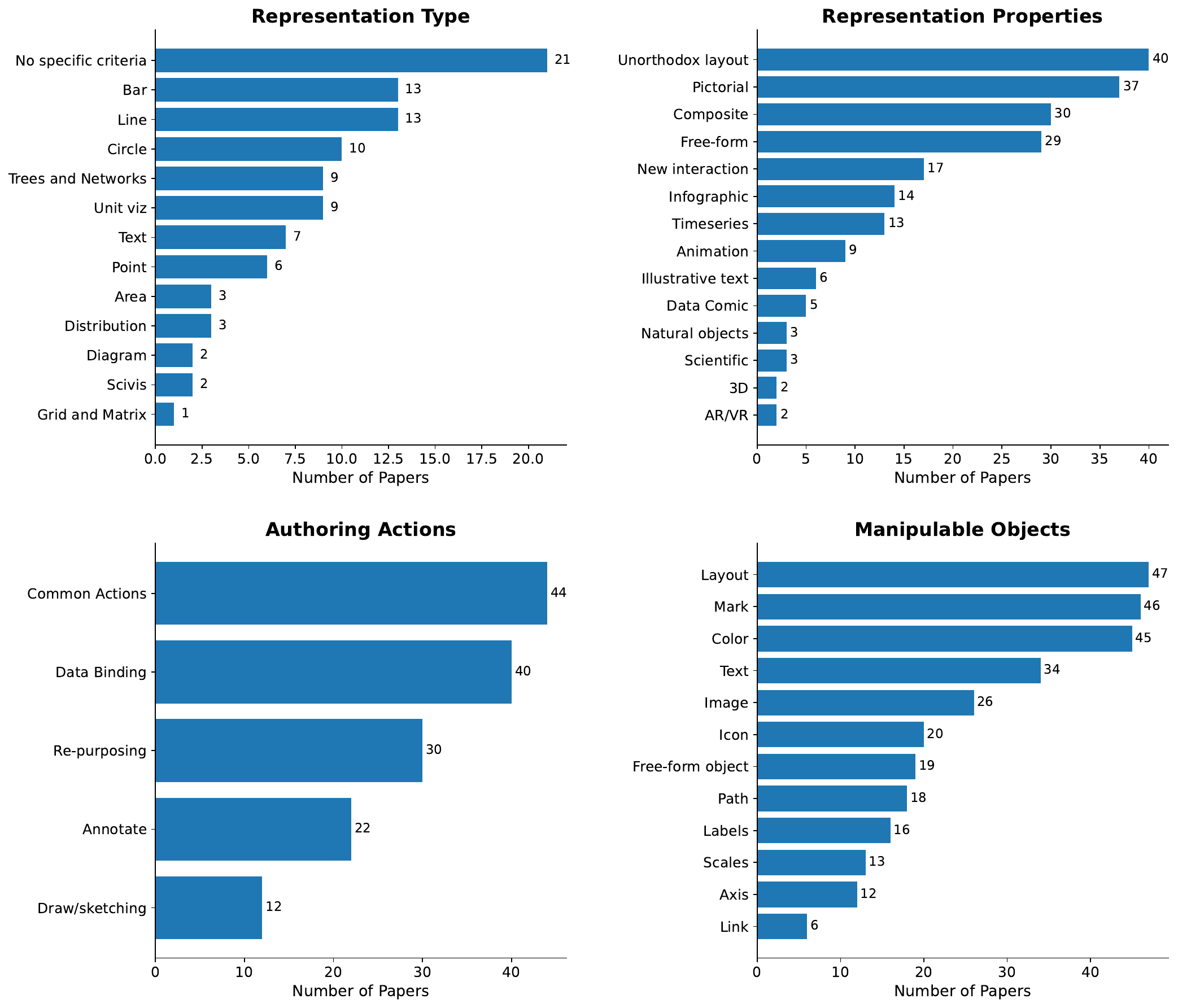}
    \caption{\textbf{Different dimensions capturing creative visualizations.} Top left shows base types of the visualizations. Top right shows properties that explain the atypical and unorthodox nature of the visualizations. Bottom left shows authoring actions that authors can perform to create the visualizations. Bottom right shows visualization objects that authors can manipulate.}
    \Description{Four horizontal bar charts in a 2×2 grid showing the number of creative representation papers by representation type, representation properties, authoring actions, and manipulable objects.}
    \label{fig:creative_rep_counts}
\end{figure}

\subsubsection{Representation Type}
We adopted the taxonomy proposed by Borkin et al.~\cite{DBLP:journals/tvcg/BorkinVBISOP13} to encode the type of data visualization that were claimed to be creative. The original taxonomy included 12 chart types (e.g., area, bar, circle, etc.).
We encountered three more chart types in our corpus, which we thus add to the list: \textit{unit visualization}, \textit{scientific visualization}, and \textit{no specific type}. Figure~\ref{fig:creative_rep_counts} (top left) presents the counts of different chart types. The most prominent type is ``No specific criteria,'' which indicates that creative charts often cannot be explained with commonly used data visualizations. This motivated us to investigate the properties of the representations more critically, which is discussed in the next dimension.

\subsubsection{Representation Properties}
\label{sec:cr_properties}
We adopted the representation properties proposed by Borkin et al.~\cite{DBLP:journals/tvcg/BorkinVBISOP13}.
The taxonomy includes properties such as \textit{pictorial, time-series, three dimensionality,} and \textit{multiplicity} that could explain the creative aspects of a representation. 
We expanded this taxonomy by adding properties encountered in the corpus. We added the following codes as properties: \textit{Free-form, AR/VR, Infographic, Scientific, Unorthodox layout, Natural objects,} and  \textit{New interaction}. Visualizations including unorthodox layouts, pictorials, composites, free-forms, and new interaction were most common (Figure~\ref{fig:creative_rep_counts} (top right)). We show examples of these properties in Figure~\ref{fig:rep_examples}.

\subsubsection{Authoring Actions}
\label{sec:cr_authoring}
We noticed creative representations can rely on author-driven design actions (e.g., drawing glyphs, mapping a data point to a mark, etc.). We coded such author-driven actions.
Our goal was to capture actions that are particularly important for a creative representation.
Thus, we combined common authoring actions (e.g., selecting data attributes and deleting specific data points)~\cite{DBLP:journals/tvcg/SatyanarayanLRH20} into a single type and added the following authoring actions to the list: \textit{Draw or Sketch, Annotate or Add Context, Data Binding or Mapping,} and \textit{Repurposing Content}. Figure~\ref{fig:creative_rep_counts} (bottom left) shows the counts of the activities. One interesting finding is that some activities from design frameworks (sketching and drawing) reappear in this theme as a direct manipulation mechanism for authoring data visualization. 
For example, SketchStory~\cite{DBLP:journals/tvcg/LeeKS13} uses free-form sketches from authors to create expressive charts.
Authors can sketch example icons which the system can replicate to match the underlying data.
The system also provides touch interactions to move, resize, and delete charts.
Similarly, TimeSplines~\cite{DBLP:journals/tvcg/OffenwangerBCT24} allows users to draw timelines of arbitrary shapes.
Users can loosely bind data to the timeline and add annotations and embellishments to the timeline. 

\subsubsection{Manipulable Objects}
This dimension captures visualization objects that authors can manipulate using the authoring actions discussed in the previous dimension.
Codes include objects such as \textit{geometric marks, glyphs, icons, color, text, legend,} etc.
The objects were inspired by the taxonomy proposed by Satyanarayan et al.~\cite{DBLP:journals/tvcg/SatyanarayanLRH20}. Layouts, marks, Colors, and Texts were the most common objects that authors can manipulate to create creative visualizations (Figure~\ref{fig:creative_rep_counts} (bottom right)).

\subsubsection{Computational Support}
This dimension captures the role of automated methods in the authoring process of a representation.
Possible values include \textit{Algorithms}, \textit{AI/ML}, and \textit{not specified}. We only account for significant automation in this dimension. Common automation methods, such as automatically detecting attribute types are not included, as they are not relevant to the topic. 43 papers out of 52 in this theme proposed some sort of computational support. Among them, 29 papers relied on algorithms only. For example, Infomages~\cite{DBLP:journals/cgf/CoelhoM20} uses heuristics and algorithms to combine images from the web and chart data for producing infographics.
Schroeder et al.~\cite{schroeder2015visualization} describe a visualization system that employs an algorithmic approach to adjust color maps through user interaction. 
The other 14 papers, authored between 2021-2026, combined algorithms with machine learning based support. For example, Cui et al.~\cite{DBLP:journals/tvcg/CuiWHWLZZ22} proposed an ML model to reuse elements from existing infographics in new ones. The MetaGlyph~\cite{DBLP:journals/tvcg/YingSDYTYW23} system uses various ML models to generate glyph-based visualizations from a spreadsheet. The remaining 9 papers did not propose or mention any significant computational support.
\begin{figure}
    \centering
    \includegraphics[width=0.9\linewidth]{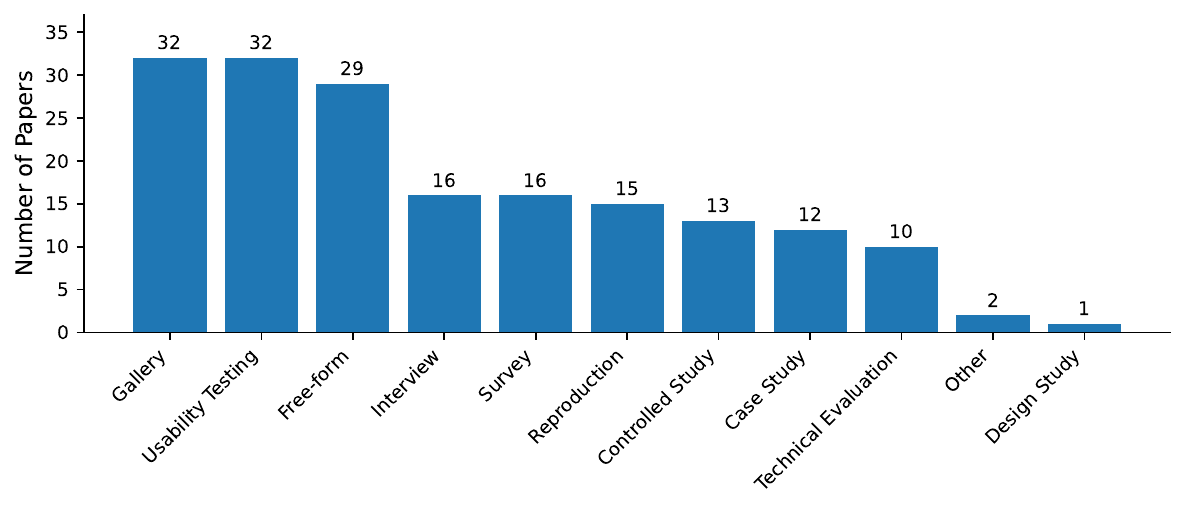}
    \caption{\textbf{Study types used to evaluate creative representations.}}
    \Description{A vertical bar chart showing the number of creative representation papers by study type, sorted from most to least common.}
    \label{fig:study_types}
\end{figure}

\subsubsection{Study Type}
Gallery and usability testing, each occurring 32 times,  were the most common user study types used to validate the tools and visualizations created in this theme (Figure~\ref{fig:study_types}). Free-form authoring to evaluate the expressiveness of the tools and techniques was also popular (29 times). This means traditional quantitative user studies are not well-suited for evaluating creative representations.

%% file: sections/4.3.cst.tex
\subsection{Creativity Support Tools (CSTs) with Visualization}
\label{sec:cst}

\begin{figure*}[t]
    \centering
    \includegraphics[width=0.92\linewidth]{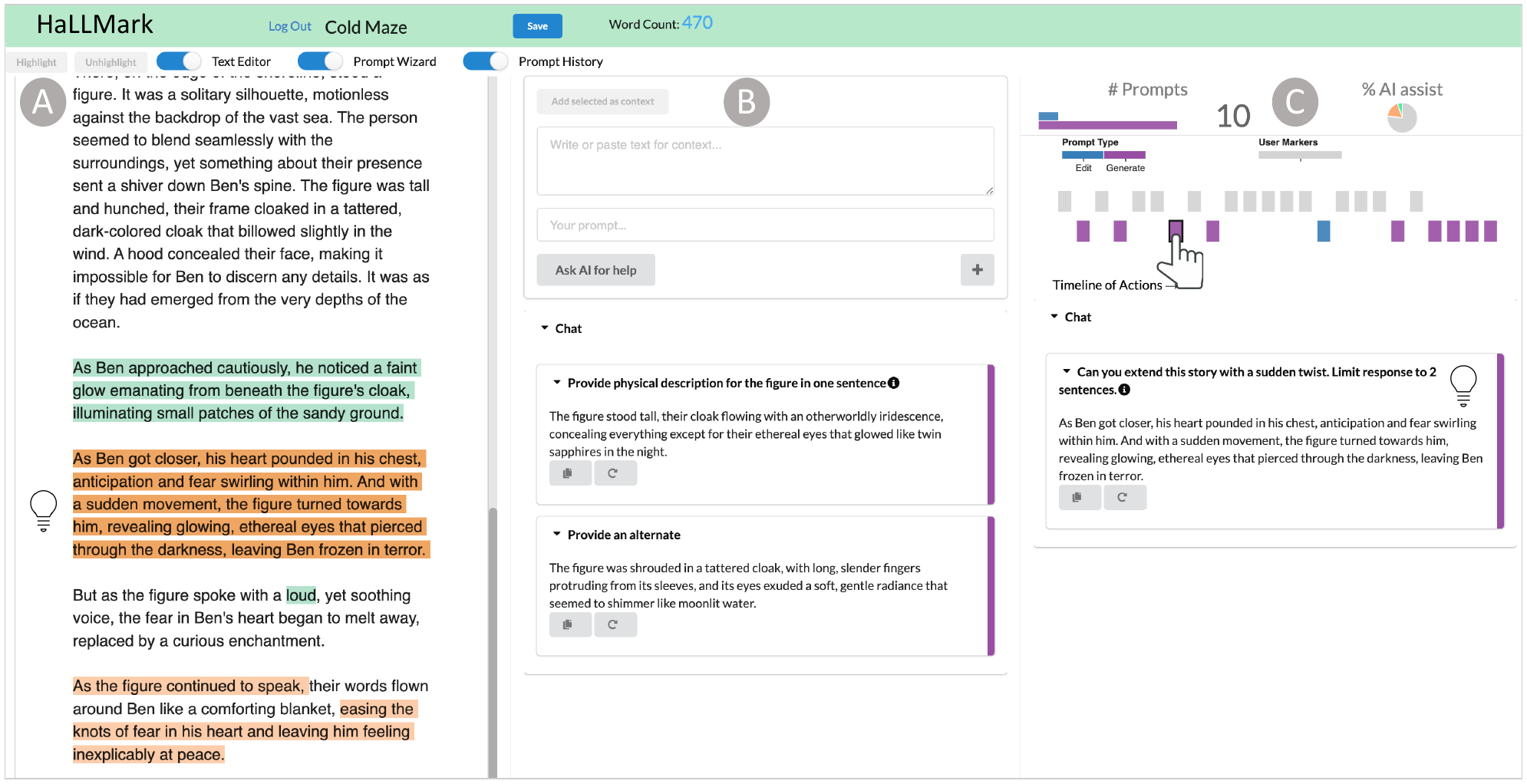}
    \caption{\textbf{HaLLMark, an example vis-enabled creativity support tool (CST).}
    The tool promotes responsible use of LLMs in writing.
    It offers a writing interface (A) with text written and influenced by the LLM highlighted with the orange and green colors, a ChatGPT like interface (B) for prompting the LLM, and a timeline visualization to trace writer-LLM interactions.
    Purple rectangles indicate prompts that were used for generating new texts whereas blue rectangles indicate prompts asking to edit an existing text.
    The writer can use this timeline to go back to any prior writing stages.}
    \Description{The Figure shows a web-based interface, where on the left side it has a text editor, in the middle some input boxes, and on the right there are some visual components.}
    \label{fig:hallmark}
\end{figure*}

This theme encompasses systems that utilize data visualization to support other creative tasks such as writing, music production, and fashion design.
Although designing visualization itself can be a creative task, this theme captures visualization in other creative tasks.~\autoref{fig:hallmark} shows HaLLMark~\cite{DBLP:conf/chi/HoqueMGSCKE24}, an example CST designed with visualization for supporting creative writing.
The tool helps writers become transparent about their usage of LLMs in accordance with existing writing policies. We discuss the dimensions of this theme in Table~\ref{tab:cst_codebook} and below.


\begin{table}[ht]
    \centering
    \caption{\textbf{Dimensions, codes, and definitions for visualization-enabled CSTs}}
    \Description{A table describing the coding dimensions, codes, and definitions for visualization-enabled creativity support tools.}
    \renewcommand{\arraystretch}{1.3}

    \begin{tabular}{m{2cm}m{6cm}m{6cm}}
    \toprule
    \textbf{Dimension} & \textbf{Code} & \textbf{Definition} \\
    \toprule

    \multirow{2}{=}{\centering Supported Task}
        & \multicolumn{2}{l}{\textit{Which creative task is supported by the CST?}} \\ 
        \dashedline{2-3}
        & Writing, video editing, music production, fashion design, landscaping, color design, etc.
        & The creative activity or domain that the system is designed to support. \\

    \hline

    \multirow{3}{=}{\centering Visualization Prominence}
        & \multicolumn{2}{l}{\textit{How prominent is visualization within the CST?}} \\ 
        \dashedline{2-3}
        & Supporting
        & Visualization is one component of a broader creative workflow and is not the primary mechanism through which the task is performed. \\

        & Central
        & Visualization substantially structures or drives the creative workflow or enables users to make sense of the creative process. \\

    \hline

    \multirow{6}{=}{\centering Visualization Function}
        & \multicolumn{2}{l}{\textit{What function does visualization serve in the creative process?}} \\ 
        \dashedline{2-3}

        & Ideation / Divergent Thinking
        & Supports the generation, exploration, or expansion of creative possibilities. \\

        & Synthesis / Convergent Thinking
        & Supports comparing, organizing, combining, refining, or selecting among creative alternatives. \\

        & Collaboration
        & Supports collaboration among people or between humans and AI during the creative process. \\

        & Data Collection
        & Supports collecting or externalizing information relevant to the creative task. \\

        & Reflection
        & Supports reviewing, interpreting, or reflecting on creative artifacts, actions, or processes. \\

    \hline

    \multirow{2}{=}{\centering Visualization Innovation}
        & \multicolumn{2}{l}{\textit{What is innovative about the use of visualization in the CST?}} \\ 
        \dashedline{2-3}

        & Open-ended
        & Describes novel visual representations, interactions, encodings, layouts, or unusual uses of visualization that support the creative task. \\



    \bottomrule
    \end{tabular}

    \label{tab:cst_codebook}
\end{table}

\subsubsection{Supported Tasks} There are 10 different types of supported tasks in this theme, indicating that visualization can support a wide variety of tasks. Design is the most popular category, encompassing 11 articles that proposed different domain specific design tasks (e.g., architecture/landscape design~\cite{DBLP:conf/infovis/BuscherSCMO00}, color-theme design~\cite{DBLP:conf/chi/ShugrinaZCFS19},  and lighting design~\cite{walch2020lightguider}). The second most popular task is writing (7 out of 28 papers)~\cite{DBLP:conf/chi/HoqueMGSCKE24, 10.1145/3301275.3302302, visual-storywriting}. 

\subsubsection{Visualization Prominence} This dimension captures the role of visualization in the CSTs. The role can be either \textit{central (visualization is the primary mechanism through which the creative activity happens)} or \textit{supporting (visualization provides information, awareness, reflection, or guidance within a broader creative activity)}. Interestingly, 78\% of papers (22/28) in this theme play \textit{central} roles in the CSTs. This means that although this set of papers primarily focused on non-visualization creative tasks, visualizations still play a key role in the tool. For example, in the FashionQ system~\cite{DBLP:conf/chi/JeonJSH21} data visualizations enable fashion designers to visually explore styles, trends, and combinations. All six papers in the supportive category used data visualization as a provenance or action tracking mechanism.

\subsubsection{Visualization Function}
This dimension further categorizes the function of data visualizations into 5 categories. The categories are similar to design activities from the first theme, although their meaning is different here. Synthesis/convergent thinking (24 out of 28 papers) and reflection (21) are the most common functions, followed by ideation/divergent thinking (17), collaboration (9), and data collection (3). 

\subsubsection{Visualization Innovation} This dimension notes an open-ended description of the innovation of the visualization. Since this set of papers primarily focused on creativity support tools, we wanted to capture the innovation around visualization in them. For example, we noticed that several papers introduced novel direct manipulation techniques to create other artifacts. Masson et al.~\cite{visual-storywriting} introduced an interface to visually map characters, locations, and events to generate stories using LLMs. 

%% file: sections/4.4.understanding_creativity.tex
\begin{table}[ht]
    \centering
    \caption{\textbf{Dimensions, codes, and definitions for Understanding Creativity (UC)}}
    \Description{A table describing the coding dimensions, codes, and definitions for studies that investigate creativity in visualization-related contexts.}
    \renewcommand{\arraystretch}{1.3}

    \begin{tabular}{lm{5cm}m{6cm}}
    \toprule
    \textbf{Dimension} & \textbf{Code} & \textbf{Definition} \\
    \toprule

    \multirow{2}{*}{\centering Domain}
        & \multicolumn{2}{l}{\textit{In what domain or context is creativity investigated?}} \\
        \dashedline{2-3}
        & Visualization design, data storytelling, music, design practice, etc.
        & The application domain or creative context in which creativity is studied. \\

    \hline

    \multirow{2}{*}{\centering Study Type}
        & \multicolumn{2}{l}{\textit{How is creativity empirically investigated?}} \\
        \dashedline{2-3}
        & Interview, Survey, Gallery, Reproduction, Free-form, Design Study,
        Technical Evaluation, Usability Testing, Case Study, Controlled Study,
        Field Study, Reading Study, Other
        & The empirical method used to investigate creativity or creative practices. \\

    \hline

    Creative Angle
        & \multicolumn{2}{l}{\textit{What aspect of creativity does the study seek to understand?}} \\ \dashedline{2-3}
        & Reflection/Creative Process, Ideation/ Inspiration, Expressive/Artistic Representation, Aesthetics/Affect/Engagement, Self-expression/Personalization, Collaboration/Co-creation
        & The aspect of creativity or creative practice investigated by the study. \\

    \hline

    \multirow{2}{*}{\centering Target Users}
        & \multicolumn{2}{l}{\textit{Whose creativity or creative practices are investigated?}} \\
        \dashedline{2-3}
        & Students, Children, Domain Experts, Visualization Experts,
        No specific criteria
        & The population whose creativity, perceptions, experiences, or creative
        practices are investigated. \\

    \hline

    Open-ended
        & Notes
        & Open-ended comments about the study or its investigation of creativity. \\

    \bottomrule
    \end{tabular}

    \label{tab:uc_codebook}
\end{table}

\subsection{Understanding Creativity}
This final theme contains research articles that aim to understand creativity, creative processes, and factors that influence creativity in data visualization. Rather than directly supporting creative activities, these studies investigate how creativity occurs and how it is shaped by factors such as inspiration, ideation, design fixation, reflection and collaboration. For example, Baigelenov et al.~\cite{DBLP:conf/chi/BaigelenovSP25} examined how visualization designers seek and use inspiration during their design process. We present the dimensions of this theme in Table~\ref{tab:uc_codebook} and below.


\subsubsection{Domain} 12 out of 17 papers focused on understanding creativity in different stages of visualization design~\cite{DBLP:conf/visualization/ParsonsSP21, DBLP:conf/chi/BaigelenovSP25}. The other 5 papers focused on understanding creativity in the CSTs from the previous theme~\cite{rau2025maico, severes2025making}.

\subsubsection{Study Type} This dimension uses similar categories as the study types that were on the creative representation theme. The main study types were interviews (8/17), surveys (7/17), and case studies (6/17). This is expected as many of these papers are trying to understand designers' experiences, practices, or perceptions rather than evaluate new visualization tools.

\subsubsection{Creative Angle}
This dimension captured the creative angles that were studied under each paper. The most common category was ``reflection (8/17)'', followed by ``ideation/inspiration (6/17)''. For example, Baigelenov et al.~\cite{DBLP:conf/chi/BaigelenovSP25} examined how visualization designers perceive and use inspiration for their designs. Since the paper asked participants to reflect on their design process, it falls under the ``reflection'' category. It also falls under the  ``ideation/inspiration'' category since the paper investigated inspiration strategies. Another example is the work by Lan et al.~\cite{Lanaffective}, which presented two crowdsourced studies to understand affective responses to infographics. This study falls under the ``aesthetics/affect/engagement'' category.  


\subsubsection{Target Users} This dimension is adopted from the design framework theme. Most papers focused on generic users with no specific criteria (10/17), followed by visualization designers (6/17) and domain experts (5/17). There were only 2 studies involving students~\cite{DBLP:journals/tvcg/BakoLBL23, iStoryline} and no papers with children. 


%% file: sections/5.interviews.tex
\section{Interview Study}

The design space helped us reflect on the current framing of creativity in visualization research.
We further conducted interviews with visualization practitioners to understand how creativity manifests in practice (e.g., in data journalism). We also recruited academic researchers who have substantial experience in this area since their experience could further strengthen the characterization of creativity. Our university's Institutional Review Board (IRB) approved the study.




\begin{table*}[tb]
    \centering
    \caption{\textbf{Participant background.} We recruited 11 participants with diverse backgrounds in data visualization and journalism.}
    \Description{A five-column table listing the ID, gender, age, profession, and experience of the 11 interview participants.}
    \label{tab:demographics}
    \begin{tabular}{@{}lllp{3.2cm}p{7.5cm}@{}}
        \toprule
        \textbf{ID} & \textbf{Gender} & \textbf{Age} & \textbf{Profession} & \textbf{Experience \& Domain Expertise} \\
        \midrule
        P1  & Female & 25-34 & Academic Researcher             & Postdoc with 8 years of research in personal data visualization\\
        P2  & Male   & 35-44 & Academic Researcher                & Associate Professor in HCI and information visualization \\
        P3 & Male   & 55+    & Academic Researcher                & Professor in information visualization, GIScience, design, and HCI \\
        P4 & Female   & 55+    & Academic Researcher                & Professor in applied creativity, HCI design, and digital technologies \\
        P5  & Female & 25-34 & Data Vis Designer         & 15 years in data science; creative data visualization; data storytelling \\
        P6  & Male   & 18-24 & Data Vis Designer               & 2+ year in data visualization and visual storytelling \\
        P7  & Female & 35-44 & Data Vis Designer \& Artist         & 15+ years in data visualization; award-winning visualization design \\
        P8  & Female & 25-34 & Data Vis Designer \& Data Journalist    & 3 years in data journalism; social media-oriented visual analytics \\
        P9  & Female & 25-34 & Data Vis Designer \& Data Journalist    & 5+ years in  visual communication; 1 year in data journalism  \\
        P10  & Male   & 35-44 & Journalist \&  Instructor & 10+ years in visual communication and interactive design \\
        P11  & Female & 45-54 & Journalist \& Instructor & 20+ years in news reporting; Journalism education \\
        \bottomrule
    \end{tabular}
\end{table*}

\subsection{Participants}

We recruited 11 participants  (Table~\ref{tab:demographics}).
Four participants are academic researchers in visualization, creativity, and HCI (P1-P4); two are journalism educators or practitioners with experience in visual communication and news reporting (P10, P11); and five are professional visualization designers (P5-P9), with P8 and P9 also working as data journalists.
Experience ranged from early-career to senior professionals, capturing both established and emerging practices.

\subsection{Procedure}

We conducted interviews over Zoom. Each session lasted approximately 60 minutes and was divided into three parts. First, we asked participants about their experience with data visualization and their broad perspective on creativity in visualization. We also asked role-specific questions about creativity: designers on their creative design process and bottlenecks; journalists on newsroom practices, decision-making around visual presentation, and constraints; and researchers on their prior research projects. Second, we presented some examples from our design space and invited participants to reflect on their relevance and limitations, again with role-tailored questions: designers on relevance to their practice; journalists on value in a newsroom context; and researchers to evaluate each category and suggest potential use cases and limitations. In the last part, participants brainstormed with the interview administrator (the first author) about how we can better define creativity in visualization, develop ecosystems, workflows, and tools to support creative processes and artifacts, the role of GenAI in this area, and potential research opportunities.
Participants who span multiple roles, such as P8 and P9, received questions drawn from both relevant sets. We encouraged participants to reflect on the discussion and reach out to us with any afterthoughts (three participants did follow up). Participants residing in the U.S. received \$20 for their time. We were not able to compensate participants outside the U.S. (3 out of 11) for organizational constraints. Their participation was voluntary. 


\subsection{Analysis}

We transcribed all interview recordings using the automatic service from Zoom.
We checked for correctness in the transcripts right after each interview when we had fresh memory from the interviews.
Two authors independently conducted open coding, generating descriptive codes drawn directly from participants’ language.
They then compared codes, resolved interpretive differences, and iteratively consolidated related codes into broader themes through thematic analysis until reaching consensus.
Emerging themes were periodically shared with the full research team for review and critical discussion.

\subsection{Findings}

We present the findings of the interviews below.

\subsubsection{Rebalancing Process and Artifact}
\label{sec:findings_process}


Participants held different views on what creativity means in visualization, but shared a common idea: the core of creativity lies in the thinking that shapes the design process, not in how ``novel'' the output appears. P10 said \textit{``of course beautiful charts are important and they are often associated with creativity, but you have to take into account the context, goals, constraints, etc. to appreciate creativity.''} Even when discussing the representation, they tied it to the context: data, goals, audience, and communicative constraints.
P7 described creativity in visualization as \textit{``making charts unique based on data and goals---not for the sake of uniqueness.''} while P9 saw it as \textit{``bringing something novel within the constraints of communication.''}
Together, they saw purpose and fit as more central to creativity than visual novelty alone.



In line with this feedback, participants appreciated that our design space captured creative representations as well as the design process for visualization and  with visualization (e.g., CSTs). They found the design space and the themes to be thought-provoking. P3, who is a visualization researcher with experience in creativity research said \textit{``There is a need/opportunity to develop a framework that helps us find the creativity in visualization research, to see where it is, how it can be supported and used, and perhaps more importantly, where it is lacking or undervalued, where it has potential.''}

Participants also noted that the creative process and the creative outcome are not always causally related.
P3 pointed out that unconventional visuals \textit{``could have been produced completely randomly---with no creative process behind them at all.''} P4 highlighted the mismatch from both sides: \textit{``you can do a process that looks like a really creative process and end up with a visualization that’s not actually that creative, or vice versa.''}
This suggests that neither process nor output alone is a reliable indicator of creativity. 

Overall, participants' feedback suggests that we need to consider the design process and outcome holistically to understand creativity. Their feedback also suggests that the design process and context are often undervalued in practice, whereas the novelty of the visualization is  considered the hallmark of creativity. As P3 said \textit{``This (design space) seems to me to take us somewhere extremely useful, because it moves the focus away from the visual artefact---something that visualization designers are of course most interested and invested in---and towards visualization knowledge, the processes, structures, and outputs that are part of that broad ecosystem.''}


\subsubsection{Design Process Is Internalized, Not Learned from Frameworks}
\label{sec:findings_internalized}

All practitioners described a stable personal workflow developed over the years: deep engagement with data, analog sketching, digital production, and iterative refinement.
Upon introducing some example frameworks from our review, practitioners indicated limited prior exposure to these structured methodologies, and none reported having employed them to support their design practice. P8 recognized doing \textit{``something like this more informally''} and P5 felt she was not \textit{``that organized.''}
P3 acknowledged that frameworks \textit{``can be helpful, but they can also be restrictive''}.
P2 was the most direct: \textit{``I've used methodology such as Five Design Sheet with students, and it has worked, but I've never used it myself because it's too prescriptive for me.''}

What emerged from these accounts is that experienced practitioners 
tend to have already formulated their own design process over time, making formal frameworks feel unnecessary or constraining. Frameworks in our design space seemed most useful for students and novices who are still building their process.

\subsubsection{Ideation Is the Bottleneck, with No Bridge 
from Exploration to Executable Ideas}
\label{sec:findings_ideation}

Ideation was identified as the most challenging phase of the creative process; not as a function of experience but as an inherent difficulty of creative work. P5, despite being an expert data visualization designer, still regarded ideation as her main bottleneck: \textit{``the creative part of  it happens at the ideation phase for me, and that's the bottleneck.''}
P7 called idea generation \textit{``the most difficult part,''} and P8 noted that \textit{``everyone struggles with ideation no matter how much experience you have, every single new dataset is a different challenge.''}
These suggest that ideation difficulty is structural rather than skill-based.

Participants also described ideation as something that requires open 
exploration and cannot be rushed or forced.
P9 valued sketching on paper because it gave her \textit{``this space with absolutely no constraints where you can really go crazy,''} yet also noted that inspiration \textit{``might come in two days,''} making strict deadlines a serious threat to creativity.
This reflects a preference for a completely free form of creativity exploration without having to commit to anything too early.
This came through in what participants wanted from tools as well: P6 explicitly called for \textit{``a comfortable vehicle''} to guide practitioners through the creative process without prescribing it, while P3 wanted tools that could \textit{``push things in my direction to get me thinking about connecting ideas and possible solutions.''}
All pointed to the same gap---support the instant transition from exploratory thinking to actionable design decisions.

\subsubsection{Creativity Needs Permission Before It Needs Support}
\label{sec:findings_permission}

Practitioners described organizational context as the main barrier to creative visualization instead of capability or tools. In journalism, constraints operate at multiple levels. P11 noted that daily news cycles leave little room for creative work, though editors would support ideas with sufficient lead time. Ethical obligations added another layer---P11 was direct: \textit{``we don't want an infographic that is misleading''} and P8 similarly described avoiding misinformation as \textit{``the biggest concern when creating data visualizations for journalism.''}

Interestingly, newsrooms came across as relatively permissive compared to other work settings: P8 noted that many of her most creative projects were newsroom-sponsored, while private clients imposed stricter requirements. P9 contrasted her data journalism role---where she had editorial support and real creative freedom---with her current position at an international organization, where specific use cases leave little room for creativity. P7 described a broader industry shift: \textit{``companies have been defaulting more towards just give me the basic bar chart---cheap and standard.''} All these accounts point to the theme---for most practitioners, what is missing is not the ability to be creative, but the organizational permission and time.

\subsubsection{Collaboration and Communication as Creative Catalysts}
\label{sec:findings_collaboration}

Participants consistently described communication and collaboration as an active part of the creative process. P6 recounted abandoning a project until a phone call with a friend, who offered encouragement and a fresh perspective, restored his motivation. P11 detailed an iterative workflow with a digital content creator: data sparked discussion, a draft followed, and repeated feedback refined the result.
P8 and P9 said editor feedback helped advance their work; P9 added that even when working alone, she deliberately sought an outside perspective.

P4 observed that involving others is usually more productive but can also introduce friction, and even seemingly solitary artists are influenced by others. Overall, participants attributed creativity in visualization to both a personal trait and a social process.

\subsubsection{The Tradeoffs of Creativity: The Learning Curve is Manageable}
\label{sec:findings_tradeoffs}

Participants acknowledged that creative representations come with a cost: novel forms require audiences to learn how to read them. 
%
%
The benefits, participants felt, generally outweigh this cost. P7 argued that strong creative work earns reader investment: \textit{``if you do it well, you will convince people to read longer---creativity entices people to actually look at the data and spend more time with it.''} P11 linked this directly to advertising revenue in journalism.
P8 found that pictorial and infographic styles aid comprehension for niche or scientific topics, while P2 emphasized creative representation's role in evoking emotion on sensitive subjects such as war or climate change. That said, participants were careful to note that the value of creative representation depends on context. As P10 put it: \textit{``I don't think anything is ever wrong to do---but is it right?''}

\subsubsection{AI as a Creative Collaborator, Not a Replacement}
\label{sec:findings_ai}

Participants were broadly skeptical of AI's role in making creative decisions. P5 put it plainly: \textit{``AI is so good at doing non-creative things.
It doesn't really understand what a good data visualization looks like.''}
In practice, most used AI for review tasks rather than generation; P8, for example, used it to check accessibility for color-blind users and screen readers.

When it came to the creative process itself, participants saw AI as 
limited. P7 compared it to a junior designer: able to produce basic work but requiring heavy guidance. At the same time, some saw 
potential in specific roles: P7 and P9 wanted AI support for early-stage inspiration, and P5 envisioned an LLM trained on designers like Tufte or Lupi that could apply their frameworks as feedback. P10 captured the underlying concern: AI can close skill gaps but may undermine \textit{``the person without the agency, knowledge, and ability''} if it takes over too much.

%% file: sections/6.definition.tex
\section{Definition: Creativity in Visualization}
\label{sec:define}
Based on previous research, design space, and interview findings, we present a tentative definition of creativity in data visualization. There is currently no consensus about creativity in data visualization, which could be beneficial for the subfield by keeping the scope of the research open. On the other hand, a working definition can help build conceptual clarity and shared vocabulary for the subfield. Similar to the definition of creativity support tools by Frich et al.~\cite{DBLP:conf/chi/FrichV0BD19}, we do not claim this definition to be exhaustive or definitive; rather, it represents an initial step toward formalizing the research subfield. We also note that although the definition is generic, its applicability beyond data visualization remains unknown at this stage.

 
\begin{tcolorbox}[colback=white, colframe=myblue,
                  colbacktitle=myblue, coltitle=black,
                  sharp corners, boxsep=3pt, left=2pt, right=2pt, top=2pt, bottom=2pt]
\textbf{Definition:}
Creativity in data visualization is a purposeful \textit{design process} shaped by \textit{divergent, convergent,} and \textit{collaborative} thinking and activities, that produces \textit{outputs} that are  \textit{original}, \textit{relevant}, and \textit{resonant}.
\end{tcolorbox}

This definition highlights the importance of both the \textit{design process} and the \textit{artifact}, which was captured in both the survey and interview study. On the design process side, the definition highlights the importance of design activities that support divergent, convergent, and collaborative thinking, situated within the real-world constraints and demands of the design context. These activities were integral to all four design space themes. Interview participants also mentioned that it is the thinking that shapes creative processes and outcomes. The inclusion of the activities puts \textbf{human creativity} in the driving seat of the design process.


The definition emphasizes that a creative output needs to be \textit{original}, as articulated by most creativity theories~\cite{boden2004creative} and our participants. The output could be a visual representation or other artifact supported by a data visualization. The keyword \textit{relevant} emphasizes that the output needs to fit the data context and  design goals. This was repeatedly emphasized by the interview participants. The final requirement is that the output should be \textit{resonant}. We borrowed the term from Meyer and Dykes~\cite{DBLP:journals/tvcg/MeyerD20} who defined it as a construct that \textit{``inspires understanding and invites action.''} We received similar feedback from the interview participants. Ultimately, the output is about the audience, and it should create a positive impact on them, inspiring them to understand it and receive the right message.

%% file: sections/7.discussion.tex
\section{Discussion}


This section presents design implications and future research directions.

\subsection{Addressing Organizational Barriers to Creative Practice}
\label{sec:discussion_permission}
 
Effective creative support for visualization requires real organizational investment. Many design frameworks in our survey rely on conditions such as co-location, day-long workshops, and sustained collaboration with domain experts (Section~\ref{sec:df_activity} and Section~\ref{sec:df_user})---all of which require institutional time and resources. Yet our interviews show that these conditions are rare: deadlines are tight, clients expect standard outputs, and organizations default to the cheapest, fastest options (Section~\ref{sec:findings_permission}). Notably, situationalist approaches, which emphasize such social and environmental factors, were also the least common among design frameworks (Section~\ref{sec:df_theory}), suggesting that research has paid less attention to the context in which creative work happens. Future research could investigate these barriers more deeply.

One way the research community can help practitioners is by building an empirical case for the value of creative visualization that organizations can resonate with. Some evidence already exists: studies in our corpus have examined audience engagement with and affective responses to expressive formats~\cite{Lanaffective}, and prior work has linked unusual visualization types and pictorial elements to memorability~\cite{DBLP:journals/tvcg/BorkinVBISOP13}. However, this evidence remains scattered across studies with different measures (Section~\ref{sec:evl}). Synthesizing these findings and translating them into outcomes that organizations value, such as reader engagement and time spent with the content, would give practitioners concrete evidence to convince their organizations to invest more time and resources.


\subsection{From Prescription to Stimulus}

Experienced practitioners found step-based design frameworks too specific (Section~\ref{sec:findings_internalized}). This is in line with the findings from Parsons~\cite{parsons}, who identified that practitioners do not follow highly systematic processes, but instead rely on situational demands and prior experience. Thus, instead of prescribing sequences of steps, future frameworks could provide \textit{stimuli} that prompt designers to think in new directions. A simpler near-term step could be to embed the taxonomy of representation types and properties from our design space (Section~\ref{sec:rq2}) into the frameworks as stimulus cards, raising questions like ``have you considered a nonlinear representation of time series or data comics?'' at key points in ideation. 

Our interviews also suggest that frameworks may be most useful for students and novices who are still developing their own design process (Section~\ref{sec:findings_internalized}). Researchers have indeed proposed several frameworks for students and children (Section~\ref{sec:df_user}), yet few studies aim to understand how these groups actually approach creative design (Section~\ref{sec:student}). In other words, frameworks are being designed for novices without a clear understanding of how novices work. Future research could first study novices' natural design processes and use these insights to design frameworks that support their developing practice without becoming restrictive.

\subsection{Supporting Ideation with AI Without Taking Over}
Ideation emerged as a common bottleneck, and the transition from open-ended exploration to concrete design decisions is poorly supported. One symptom is the gap between analog and digital: pen-and-paper allows unconstrained thinking, whereas digital tools force early commitments to chart types and layouts (Section~\ref{sec:findings_ideation}). Prior work has also linked authoring tools that do not support exploring alternative designs to design fixation~\cite{DBLP:conf/visualization/ParsonsSP21}.

Large language models (LLMs) and vision-language models (VLMs) offer opportunities to bridge this gap. Our design space shows that sketching serves both as an ideation activity in design frameworks and as an authoring action in tools for creative representations, yet the two are rarely connected (Section~\ref{sec:gap}). Future tools could use VLMs to interpret rough sketches as design intent and generate multiple data-bound alternatives from a single loose drawing, along the lines of DataGarden~\cite{offenwanger2024datagarden}, rather than requiring upfront encoding decisions. AI could also serve as a source of stimuli during divergent thinking, generating relevant analogies, provocative design questions, or precedents matched to a dataset's structure and domain, provocative design questions, or precedents matched to a dataset's structure and domain, and could participate in discussion and critique as a simulated team member~\cite{DBLP:conf/uist/ParkOCMLB23}. Later in the process, once multiple candidates exist, lightweight interfaces could help practitioners compare alternatives across factors such as data fidelity, audience needs, and platform constraints, supporting more deliberate convergence.

However, such support must respect the concerns our participants raised. Participants were skeptical of AI making creative decisions, compared it to a junior designer requiring heavy guidance, and worried that relying on AI could undermine practitioners' agency and skills (Section~\ref{sec:findings_ai}). At the same time, some wanted AI support for early-stage inspiration and feedback. These views suggest that AI should expand the space of possibilities rather than select among them: it can propose alternatives, surface precedents, and offer critique, while leaving creative judgment to the designer. This aligns with our definition of creativity (Section~\ref{sec:define}), which places human divergent, convergent, and collaborative thinking at the center of the design process.

\subsection{Tools for Data Arts}
\begin{figure}
    \centering
    \includegraphics[width=0.95\linewidth]{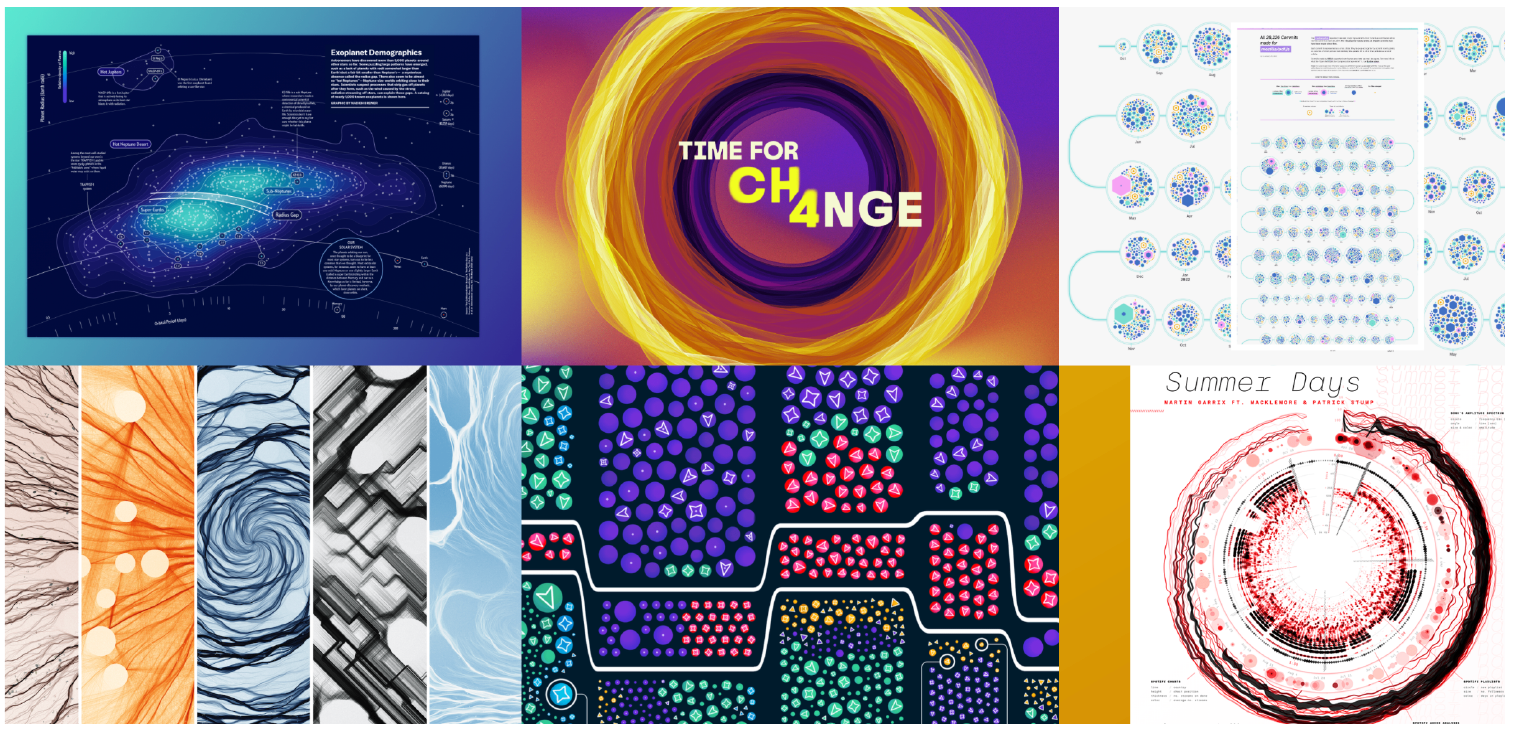}
    \caption{\textbf{A snapshot of data arts by Nadieh Bremer~\cite{bremer2025chart}.}
    Such artistic visualizations were rare in our corpus, indicating a gap in literature. Future research can pursue understanding such artistic representations and develop AI-powered authoring tools to lower barriers for this kind of representations.}
    \Description{This is a gallery of different colorful and unorthodox data visualizations.}
    \label{fig:data_arts}
\end{figure}

Another promising direction is designing \textbf{specialized authoring tools for data arts}.
While the representations discussed in the design space are largely unorthodox (Section~\ref{sec:cr_properties}), they still lack artistic expression that could be found in the brush of a data or visual artist such as Nadieh Bremer~\cite{bremer2025chart} (\autoref{fig:data_arts}).
One possible reason behind this absence may be that few tools were designed with expert visual artists as target users. 
Future research can overcome this shortcoming by engaging with visual artists to understand their practices and the type of support that could be useful to them. 
Technical works can model artist interaction into digital forms, develop rendering techniques, and design tools for artistic expression.

\subsection{Evaluating Creativity in Visualization}
Devising \textbf{new methods for evaluating creativity} is a promising research direction.
The absence of standardized evaluation methods (Section~\ref{sec:evl}) suggests a need for clearer guidelines in assessing creative contributions to visualization design.
We identified a few guidelines from our analysis that can work as seeds for future evaluation criteria:


\begin{itemize}
    \item For quantitative studies, utilize metrics (number of unique solutions explored, serendipitous solutions, satisfaction with solutions, etc.) proposed by Scholtz~\cite{DBLP:conf/ieeevast/Scholtz06}.
    Researchers can also consider solutions such as the Torrance Tests of Creative Thinking ~\cite{torrance1966torrance}.
    
    \item For qualitative studies, consider evaluations such as gallery, free-form, and design activities that can show creative exploration.
    These methods are common for measuring creative tasks in HCI~\cite{DBLP:conf/chi/FrichV0BD19}.
    
    \item Consider directly asking about creativity while gathering user feedback through interviews or surveys.
    We noticed that few reported explicitly asking participants about creativity, resulting in missed opportunities to evaluate creativity.
    
\end{itemize}



\subsection{Limitations}


This work faces some methodological and conceptual limitations.
For one, our keyword-based screening approach may have created systematic exclusions.
Papers describing genuinely innovative techniques such as treemaps~\cite{DBLP:journals/ivs/ShneidermanDSW12} or parallel coordinates do not appear in our corpus because their authors framed contributions as technical advances rather than creative endeavors.
This selection bias means we analyzed papers that \textit{claim} creativity rather than papers that \textit{demonstrate} it.
The resulting corpus may thus overrepresent explicit creativity rhetoric while missing implicit creative practices. We believe the interviews were particularly important to overcome this limitation to some extent. We were able to capture broader perspectives from the academic and practitioner communities through the interviews.



The fundamental challenge remains with the concept of creativity. As mentioned early in the paper, creativity is an abstract concept and is often dubbed an enigma by researchers~\cite{sternberg1999handbook}. Even after our best efforts, we acknowledge that some explanations in the paper could remain abstract or confusing.

%% file: sections/8.conclusion.tex
\section{Conclusion}

We presented two complementary studies to characterize creativity in data visualization.
A systematic review of 104 papers yielded a design space of four themes, showing that research supports creativity through both design processes and artifacts, yet these strands remain largely disconnected. Interviews with 11 practitioners and researchers revealed that, the final artifact is often treated as the primary indicator of creativity in practice, whereas the design process remains undervalued; participants also identified ideation as a common bottleneck and organizational constraints as a major barrier to creative work. Synthesizing both studies, we proposed a working definition of creativity in data visualization and outlined design implications and research directions. We hope this work helps the community better support both creative processes and artifacts in the age of AI-assisted visualization design.
